\documentclass[twocolumn]{aastex701}
\usepackage{amsmath}

\newcommand{\ha}{H$\alpha$}
\newcommand{\hb}{H$\beta$}

\newcommand{\oiii}{[O\,\textsc{iii}]}
\newcommand{\oiiiw}{[O\,\textsc{iii}]~$\lambda$5006.843}
\newcommand{\nii}{[N\,\textsc{ii}]}

\newcommand{\niibw}{[N\,\textsc{ii}]~$\lambda$6583.460}
\newcommand{\niiabw}{[N\,\textsc{ii}]~$\lambda$6548.050,6583.460}
\newcommand{\sii}{[S\,\textsc{ii}]}
\newcommand{\siiabw}{[S\,\textsc{ii}]~$\lambda$6716.440,6730.810}

\graphicspath{{./}{figures/}}

\begin{document}

\title{The Small-scale Structures in the Wind of Messier 82}

\author[0000-0002-6137-0422]{Yucheng Guo}
\affiliation{School of Earth \& Space Exploration, Arizona State University, 781 Terrace Mall, Tempe, AZ 85287, USA}
\email[show]{yuchengg@asu.edu}  

\author{Timothy Heckman} 
\affiliation{School of Earth \& Space Exploration, Arizona State University, 781 Terrace Mall, Tempe, AZ 85287, USA}
\affiliation{Department of Physics \& Astronomy, Johns Hopkins University, Bloomberg Centre, 3400 N. Charles Street, Baltimore, MD 21218, USA}
\email{theckma1@jhu.edu}

\author[0000-0002-2724-8298]{Sanchayeeta Borthakur}
\affiliation{School of Earth \& Space Exploration, Arizona State University, 781 Terrace Mall, Tempe, AZ 85287, USA}
\email{sborthak@asu.edu}

\author[0000-0002-1333-147X]{Peixin Zhu}
\affiliation{Center for Astrophysics $|$ Harvard \& Smithsonian, 60 Garden Street, Cambridge, MA 02138, USA}
\email{peixin.zhu@cfa.harvard.edu}

\author[orcid=0000-0003-3351-0878]{Rosalia O'Brien}
\affiliation{Department of Astronomy, University of Maryland, College Park, MD 20742, USA}
\affiliation{Astrophysics Science Division, Code 660, NASA Goddard Space Flight Center, 8800 Greenbelt Rd., Greenbelt, MD 20771, USA}
\affiliation{Center for Research and Exploration in Space Science and Technology, NASA/GSFC, Greenbelt, MD 20771 USA}
\email{rosalia.d.obrien@nasa.gov}

\author[0000-0002-6620-7421]{Ralph S. Sutherland}
\affiliation{Research School of Astronomy and Astrophysics, Australian National University, Australia}
\email{Ralph.Sutherland@anu.edu.au}

\author[0000-0001-8152-3943]{Lisa J. Kewley}
\affiliation{Center for Astrophysics $|$ Harvard \& Smithsonian, 60 Garden Street, Cambridge, MA 02138, USA}
\email{lisa.kewley@cfa.harvard.edu}

\author[orcid=0000-0003-3498-2973]{Lee Armus}
\affiliation{IPAC, California Institute of Technology, 1200 East California Boulevard, Pasadena, CA 91125, USA}
\email{lee@ipac.caltech.edu}

\author[0000-0003-0645-5260]{D. B. Fisher}
\affiliation{Centre for Astrophysics and Supercomputing, Swinburne University of Technology, P.O. Box 218, Hawthorn, VIC 3122, Australia}
\affiliation{ARC Centre of Excellence for All-Sky Astrophysics in 3 Dimensions (ASTRO 3D), Australia}
\email{dfisher@swin.edu.au}

\author[0000-0002-2644-0077]{Sebastian Lopez}
\affiliation{Department of Astronomy, The Ohio State University, 140 W. 18th Ave., Columbus, OH 43210, USA}
\affiliation{Center for Cosmology and AstroParticle Physics, The Ohio State University, 191 W. Woodruff Ave., Columbus, OH 43210, USA}
\email{lopez.764@buckeyemail.osu.edu}

\author[orcid=0000-0002-4271-0364]{Brant E. Robertson}
\affiliation{Department of Astronomy and Astrophysics, University of California, Santa Cruz, 1156 High Street, Santa Cruz, CA 95064, USA}
\email{brant@ucsc.edu}

\author[0000-0001-9735-7484]{Evan E. Schneider}
\affiliation{Department of Physics and Astronomy, University of Pittsburgh, 3941 O’Hara St. Pittsburgh, PA 15260}
\email{eschneider@pitt.edu}

\author{Patrick L. Shopbell}
\affiliation{Astronomy Department, California Institute of Technology, MC 249-17, Pasadena, CA 91125}
\email{pls@astro.caltech.edu}

\begin{abstract}
Small-scale multiphase structure plays a central role in galactic-wind evolution, yet the parsec-scale morphology and excitation of the warm ionized gas remain poorly constrained.
We present deep \emph{HST} narrow-band imaging of the southern wind of Messier~82 (M82) in \ha, \oiii, \sii, and \nii, designed to resolve the warm ionized phase on parsec scales. The \ha\ emission is detected to $\approx2.1$~kpc above the disk, while the fainter emission lines are detected over smaller radial extents, with \oiii\ reaching $\approx1.5$~kpc. We develop a filament-finding pipeline for the \ha\ image and construct a quantitative catalogue of the filamentary structures. The wind forms a highly connected network of strands and knots, dominated by compact filaments with typical projected widths of $\simeq5.3$~pc and lengths of $\simeq9.5$~pc.
Both the projected covering fraction and the line-flux contribution of the filamentary component decline with height, showing that the outer wind becomes increasingly dominated by diffuse emission. 
Optical line-ratio diagnostics indicate that the warm ionized gas occupies an intermediate excitation regime: photoionization by the central starburst can energetically power the observed \ha\ luminosity, while the systematic separation between filamentary and diffuse emission, together with the evolution of the line ratios with vertical distance, suggests an increasing contribution from shocks or other similar heating in the diffuse outer wind.
These results show that separating filamentary and diffuse emission in high-resolution imaging provides a powerful way to connect the morphology, excitation, and multiphase structure of galactic winds.
\end{abstract}

\keywords{\uat{Galaxies}{573} --- \uat{Galaxy winds}{626} --- \uat{Starburst galaxies}{1570} --- \uat{Stellar feedback}{1602} }



\section{Introduction} 

Galactic winds are a fundamental consequence of stellar feedback \citep{thompson24}. By transporting mass, momentum, energy, and metals out of galactic disks, they reshape the interstellar medium (ISM) and the circumgalactic medium, and help regulate the long-term fuel supply for star formation \citep{hopkins11,schaye15}. Galactic-scale winds are observed across a wide range of star-forming systems and are frequently described as bipolar or biconical structures oriented along the minor axis \citep{guo23}, reflecting the preferential escape of accelerated gas through low-density channels above and below the disk.

Multi-wavelength observations have established that winds are intrinsically multiphase, spanning hot X-ray--emitting plasma ($T \gtrsim 10^{6}$~K), warm ionized gas ($T \sim 10^{4}$--$10^{5}$~K) traced by optical/UV lines, cool atomic gas ($T \sim 10^{2}$--$10^{4}$~K), and cold molecular gas ($T \lesssim 10^{2}$~K) \citep{veilleux20}. 
Yet, despite the rich observations, key aspects of wind launching and propagation remain poorly understood. In particular, it is still unclear how the warm ionized phase is produced, heated/ionized, and accelerated to the observed velocities of hundreds to $\gtrsim10^{3}$~km~s$^{-1}$, and how this phase exchanges mass and energy with the surrounding hot wind fluid \citep{thompson24}.

Theoretical models struggle to explain how warm gas survives and is accelerated within a fast, tenuous hot wind. In the classical picture of ram-pressure acceleration, dense clouds are accelerated by the hot flow, but hydrodynamic instabilities and shocks can disrupt them before they reach the observed outflow velocities \citep[e.g.][]{klein94,scannapieco15,schneider17}. This has motivated extensive work on radiative turbulent mixing, cloud growth, and additional physics such as thermal conduction, magnetic fields, and cosmic rays \citep[e.g.][]{gronke18,sparre19,sparre20,fielding20,abruzzo22}. 
In these models, the fate of the gas clouds depends sensitively on the cloud size, column density, internal density structure, and the efficiency of cooling in the mixed gas. 
A recurring conclusion from these studies is that the small-scale structure is a controlling factor in wind evolution. The cloud size and density distribution regulate survival times, covering fractions, momentum coupling, and mass exchange between phases, yet the predicted morphology remains sensitive to numerical resolution and unresolved microphysics \citep[e.g.][]{thompson24,gronke26}. Direct measurements of cloud and filament sizes are therefore essential for connecting observations to simulations. Such parsec-scale constraints are currently feasible only in a few nearby, high-surface-brightness winds.

Messier~82 (M82), one of the nearest starburst galaxies, provides an unparalleled laboratory for addressing these questions. Its proximity ($D\simeq3.6$--3.8~Mpc) and near edge-on orientation ($i\approx80^\circ$) yield a uniquely favourable view of the bipolar nebula emerging from the stellar midplane \citep{lynds63}. 
M82 resides in the dynamically complex environment of the M81 triplet and is embedded in an extended halo of {H~\small{I}} streams and filaments \citep{deblock18}. This large-scale neutral reservoir implies that the outflow is not propagating into an empty halo, but instead interacts with pre-existing circumgalactic and intra-group gas. Such an environment may help confine the wind, promote cooling and mixing, and generate cloud--wind or shell-like shocks at the interfaces between the outflow and the surrounding material.

M82's wind has been characterised across nearly all major gas and dust phases. Cold atomic and molecular gas trace the massive entrained component of the wind \citep[e.g.][]{taylor01,walter02,leroy15,chisholm16,deblock18,martini18,yoshida19,krieger21,levy23,fisher25,cronin26}. Recent JWST imaging has further revealed a spectacular network of PAH-emitting filaments in the inner several kiloparsecs of the wind, showing that dust-bearing cool gas survives far above the disk and remains closely connected to the cold molecular phase \citep{fisher25,cronin26,lopez26}. 
The warmer phase is traced by optical emission lines such as \ha\ and \nii\ \citep[e.g.][]{shopbell98,devine99,mutchler07,lokhorst22,xu23,lopez25}, while the hot wind fluid is seen in soft X-ray emission \citep[e.g.][]{lehnert99,strickland07,lopez20}. In the standard picture, the X-ray-emitting plasma represents the hot, volume-filling wind, whereas the optical filaments arise in cooler material at cloud surfaces, mixing layers, or shocked interfaces. 
Together, these observations motivate a picture in which the M82 wind is a multiphase flow interacting with both entrained cool clouds and the surrounding environment.

In this work, we present deep HST/WFC3 narrow-band imaging of the south-eastern (approaching) half of the M82 wind, targeting \ha, \oiii, and \sii, with a broad-band image for continuum subtraction. We develop a customised filament-detection pipeline and construct a quantitative filament catalogue. Combining multiple emission lines, we aim to provide empirical constraints on the characteristic scales and excitation of the warm ionized phase, and to connect these constraints to emerging theoretical pictures of multiphase wind structure.

\section{Data} 

\begin{figure}
    \centering
    \includegraphics[width=1\columnwidth]{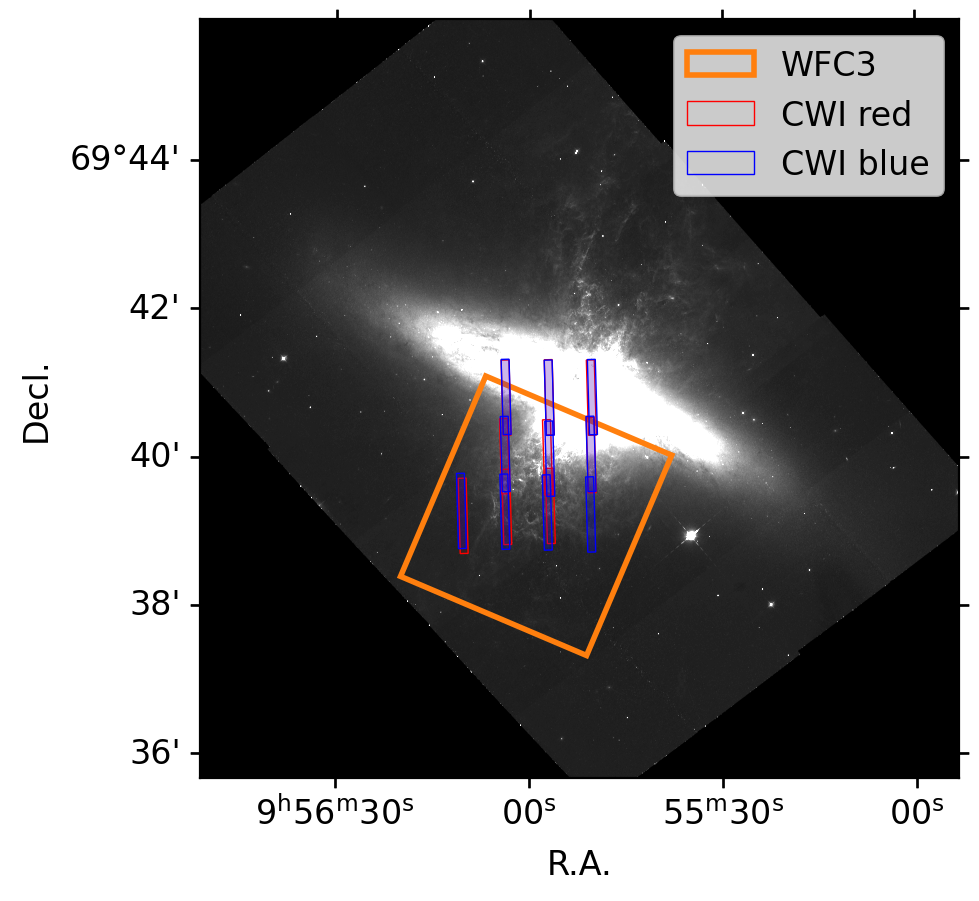}
    \caption{Footprint of the \emph{HST} WFC3/UVIS observations (yellow) overlaid on the archival \emph{HST} ACS F555W mosaic of M82 \citep{mutchler07}. The regions covered by the CWI observations are also indicated.}
    \label{fig_field}
\end{figure}

Our \emph{HST} WFC3/UVIS imaging (Program ID: 17542, PI: T.~Heckman,  \dataset[doi: 10.17909/kd75-ma21]{http://dx.doi.org/10.17909/kd75-ma21}) covers the south-eastern half of the galactic wind in M82. Both the gas kinematics and the morphology of the M82 disk indicate that the south-eastern region corresponds to the approaching side of the outflow from our vantage point, and is therefore less affected by dust obscuration within the disk. This makes it particularly well suited for studying the wind structure. The relative location of the WFC3 field of view with respect to M82 is shown in Figure~\ref{fig_field}.  
The bottom of the WFC3 field of view is approximately 0.28~kpc away from the disk of M82.

The observations were obtained through four filters: the narrow-band filters F502N, F656N, and F673N, and the broad-band filter F606W. The narrow-band imaging is designed to isolate key optical emission lines tracing ionized gas in the wind, with F502N targeting \oiiiw, F656N targeting \ha\ (with minor contamination from the \niiabw\ doublet), and F673N targeting the \siiabw\ doublet. The broad-band F606W image is used to characterise the underlying stellar continuum. A summary of the \emph{HST} observations is provided in Table~\ref{tab_HST_obs}.

\begin{figure}
    \centering
    \includegraphics[width=1\columnwidth]{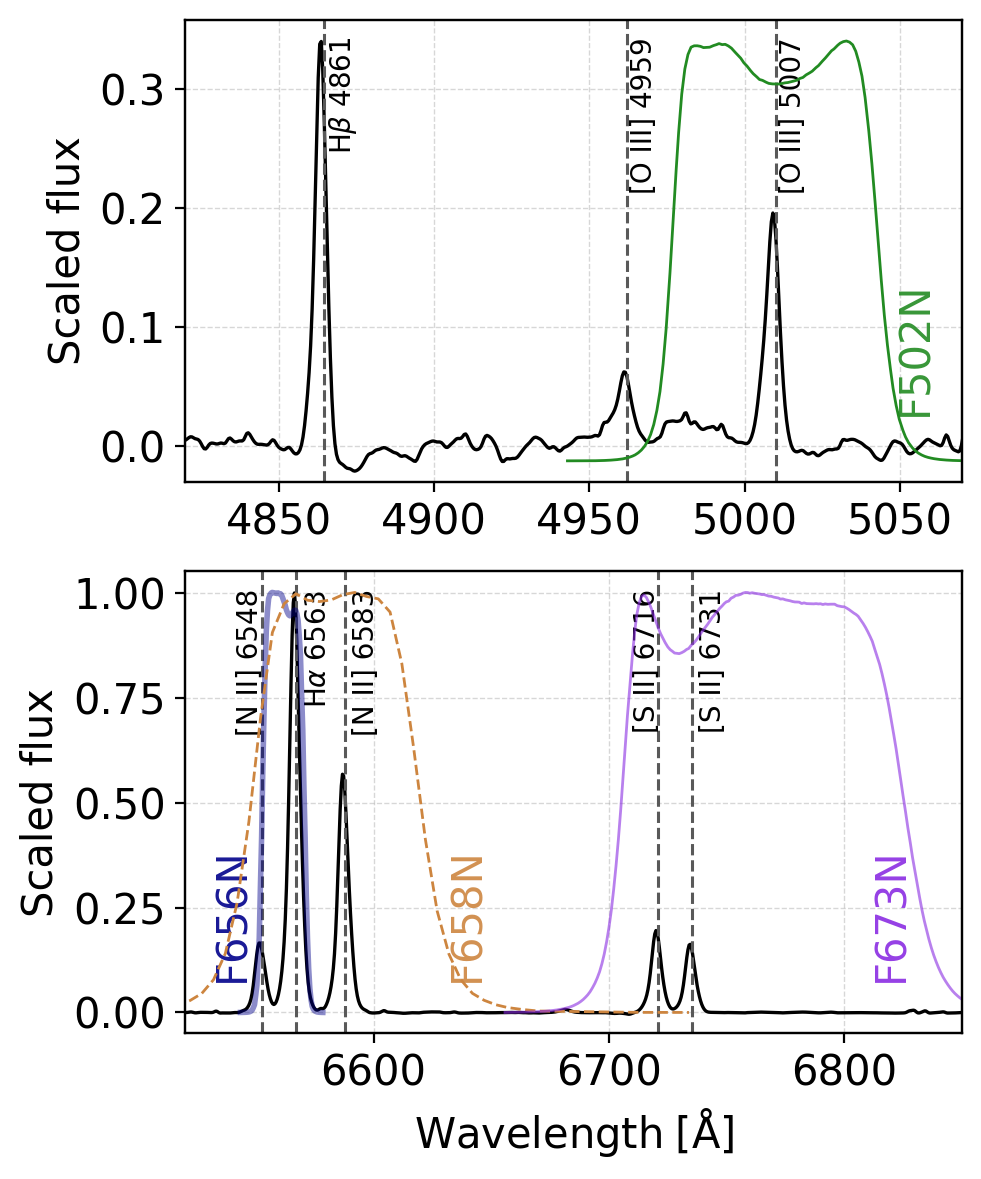}
    \caption{Median CWI spectra extracted from the region overlapping with the WFC3 field of view. The spectra are not flux calibrated; therefore, the ordinate is shown in arbitrary units. The transmission curves of the WFC3 narrow-band filters, together with the F658N filter used by \citet{mutchler07}, are overplotted.}
    \label{fig_filter}
\end{figure}

The \emph{HST} imaging data are supplemented by integral-field spectroscopic observations obtained with the Palomar Cosmic Web Imager (CWI) during 2014--2015 (PI: J.~Rich). The CWI observations partially overlap with the WFC3 field of view, as shown in Figure~\ref{fig_field}, and cover an approximately $7''\times64''$ region on the sky. Both the Blue and Red gratings were used, providing spectral coverage over the wavelength ranges 4800--5200~\AA\ and 6300--7000~\AA, respectively. Typical individual exposure times are 300--600~s.
Median CWI spectra extracted from the overlapping region are shown in Figure~\ref{fig_filter}, together with the transmission curves of the relevant WFC3 narrow-band filters. The CWI spectra are not flux calibrated, so the ordinate is shown in arbitrary units.
Because the southern region is the approaching side of the outflow, the emission lines are slightly blueshifted relative to their systemic wavelengths. This velocity shift is small compared with the filter bandpasses, and does not move the target emission lines out of the corresponding narrow-band filters (Figure~\ref{fig_filter}).

\begin{table*}
    \caption{\emph{HST} WFC3 observations (PID=17542).}
    \label{tab_hstobs}
    \centering
    \begin{tabular}{lcccc}
        \hline
        Filter & Exposure Time & Central Wavelength & Width & Emission line \\
         &  (s) & (\AA) & (\AA) & \\
        \hline
        F656N & 13792.0 & 6561.4 & 17.6 & \ha   \\
        F673N & 27584.0 & 6765.9 & 117.8 & \sii  \\
        F502N & 27584.0 & 5009.6 & 65.3 & \oiii \\
        F606W & 2379.0  & 5888.4 & 2189.2 & Continuum  \\   
        \hline
    \end{tabular}
\label{tab_HST_obs}
\end{table*}

\begin{figure*}
    \centering
    \includegraphics{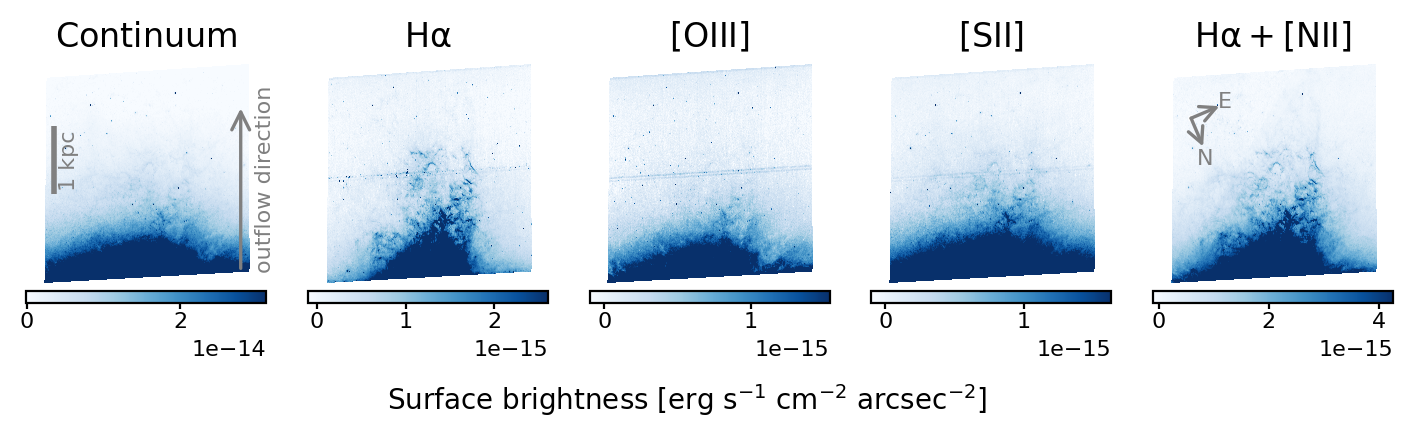}
    \caption{All \emph{HST} broad- and narrow-band images used in this work. The left four panels show our WFC3 imaging in F606W, F656N, F502N, F673N. The right-hand panel shows the archival ACS F658N image from \citet{mutchler07}, whose bandpass includes both \ha\ and \nii\ emission lines. The image stretch is adjusted separately for each panel.}
    \label{fig_original_images}
\end{figure*}


\subsection{HST data calibrations}

The \emph{HST} WFC3/UVIS imaging was processed using the calibrated \texttt{\_flc} files produced by the standard \emph{HST} pipeline, which include bias subtraction, dark correction, flat-fielding, and pixel-based correction for charge transfer inefficiency. For each filter, individual exposures were combined using \texttt{AstroDrizzle} within the \texttt{DrizzlePac} package \citep{fruchter10,hoffmann21}.

Given the low surface-brightness nature of the wind and the limited number of compact sources in the field, we did not apply additional relative alignment with \texttt{TweakReg}. 
The astrometric consistency between exposures is better than the pixel RMS, and thus negligible for the surface-brightness analysis presented here. Cosmic rays were rejected using the \texttt{minmed} combination algorithm with conservative rejection thresholds. We adopted a square drizzle kernel and disabled sky subtraction during the drizzling step in order to avoid oversubtraction in the presence of extended wind emission. The final drizzled images retain the calibrated count-rate units (electrons~s$^{-1}$).
All images were reprojected onto a common astrometric grid to enable pixel-by-pixel comparison and line-ratio measurements. 

Because the extraplanar wind emission fills a substantial fraction of the field of view, standard global background subtraction can bias measurements of faint diffuse structures. We therefore adopted a two-stage background estimation procedure. First, we applied a conservative mask excluding the bright outflow region. We then estimated the large-scale background using a two-dimensional median filtering approach with sigma clipping to suppress residual emission and detector artifacts. 
Statistically significant ($>3\sigma$) positive residual structures were identified and adaptively masked. 
A refined median background value was subsequently recomputed using the expanded mask and subtracted from the science images. This iterative procedure minimizes oversubtraction while preserving low surface-brightness wind emission.

The final drizzled images were converted to physical flux units using the \texttt{PHOTFLAM} keyword provided in the image headers, which converts count rate to flux density in units of erg~s$^{-1}$~cm$^{-2}$~\AA$^{-1}$. The integrated line fluxes were later obtained by multiplying by the \texttt{PHOTBW} keyword from the image headers. 
The reduced WFC3 broad and narrow band images are shown in Figure~\ref{fig_original_images}.

\subsection{Continuum subtraction} 
\label{subsec_cont_sub}

In this work, the narrow-band images are used to map emission-line fluxes, while the F606W image is employed to estimate and subtract the continuum contribution from the narrow-band data. 
Because the F606W bandpass overlaps several strong nebular lines present in the wind spectrum, we first correct the F606W image for emission-line contamination before using it as a pure continuum tracer.

In practice, we correct the F606W flux density $f_\lambda^{606}$ by subtracting the predicted line contributions from H$\alpha$ and \nii\ (F656N), \oiii\ (F502N), and \sii\ (F673N), yielding a ``clean'' continuum image

\begin{equation}
\begin{aligned}
f_{\lambda,\mathrm{cont}}^{606} \;=\;& f_\lambda^{606}
\;-\; C_{1}\left(\frac{W_{656}}{W_{606}}\right) f_\lambda^{656}
\;-\; C_{2}\left(\frac{W_{502}}{W_{606}}\right) f_\lambda^{502} \\
&-\; \left(\frac{W_{673}}{W_{606}}\right) f_\lambda^{673}.
\end{aligned}
\end{equation}
where $f_\lambda^{656}$, $f_\lambda^{502}$, and $f_\lambda^{673}$ are the narrow-band flux-density maps (already background-subtracted and calibrated), and $W$ denotes the corresponding rectangular widths.

The coefficients $C_1$ and $C_2$ correct for emission-line complexes that are not uniquely captured by a single narrow-band filter but nevertheless contribute to the broad F606W passband. 
For \oiii, we account for both lines of the doublet by adopting the theoretical ratio $F_{5007}/F_{4959}\simeq 3$ \citep[e.g.][]{osterbrock06}, which implies $C_2 = 1 + \frac{1}{3}$. 
For the Balmer contribution traced by $f_\lambda^{656}$, we include H$\alpha$ and H$\beta$ emission that falls within F606W by assuming Case~B recombination with $F_{\mathrm{H}\alpha}/F_{\mathrm{H}\beta}\simeq 3$ \citep[e.g.][]{hummer87,osterbrock06}. 
In addition, we correct the \nii\ contribution using a representative ratio H$\alpha$/\nii\ $\simeq 0.95$ measured from the overlapping CWI spectroscopy (Figure~\ref{fig_filter}), and we propagate this correction consistently into the broadband contamination term.

\begin{figure*}
    \centering
    \includegraphics[width=\textwidth,height=0.9\textheight,keepaspectratio]{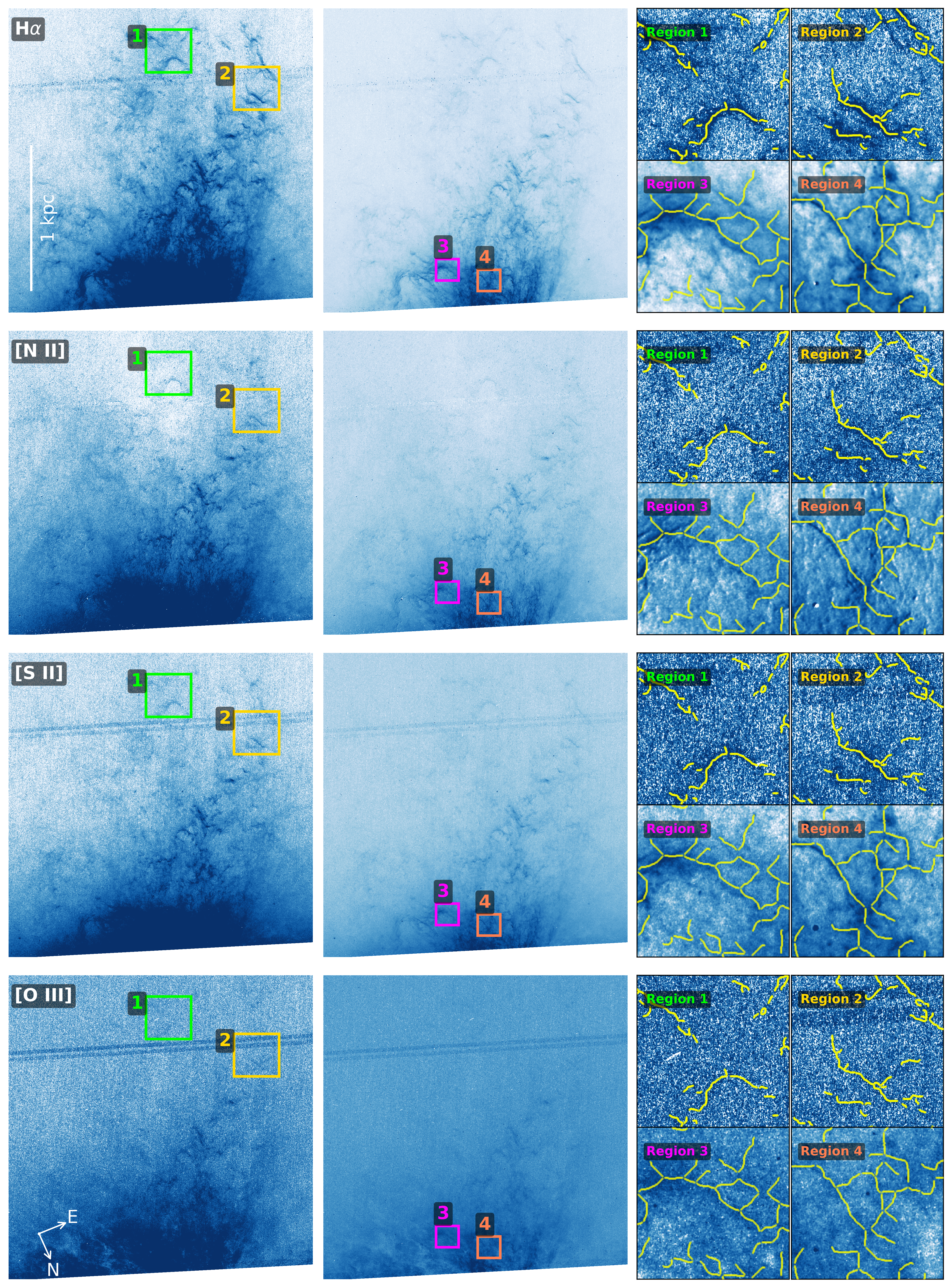}
    \caption{Multi-line view of the M82 outflow and its substructures. From top to bottom, the rows show the continuum-subtracted \ha, \nii, \sii, and \oiii\ emission-line maps. 
    For each tracer, the first two columns show the same cropped field, displayed with different stretches to better visualise the outer and inner parts of the wind, respectively. Four representative regions are marked by coloured boxes: Regions~1 and~2 sample the upper outflow, while Regions~3 and~4 lie closer to the base of the wind. The third column shows zoom-in views of these four regions, with yellow lines marking the identified filamentary structures. The same regions are shown for all tracers, illustrating the broadly consistent morphology of the ionized filaments across the tracers.    
}
    \label{fig_line_images}
\end{figure*}

The resulting cleaned continuum map, $f_{\lambda,\mathrm{cont}}^{606}$, is then used to remove the stellar continuum from the narrow-band images, yielding continuum-subtracted emission-line maps. 
The first two columns of Figure~\ref{fig_line_images} show the emission-line maps displayed with different stretches to better visualize the outer and inner parts of the wind, respectively.
This procedure assumes that the stellar continuum is approximately flat across each narrow-band passband, which is well justified given the very small wavelength intervals involved and the smoothly varying optical continua expected for the stellar populations dominating the F606W light. We have further verified that modest variations in the adopted coefficients $C_1$ and $C_2$ do not qualitatively change the continuum-subtracted maps or any subsequent measurements, demonstrating that our results are robust to the precise choice of these correction factors within a plausible range.

\subsection{Archival data} 

To include an additional diagnostic line constraint, we also make use of the \emph{HST}/ACS narrow-band F658N imaging presented by \citet{mutchler07}.
These observations were obtained as part of \emph{HST} program 10776 (PI: M.~Mountain), including broad-band imaging in F435W, F555W, and F814W, together with narrow-band imaging in F658N. 
The F658N imaging was labelled H$\alpha$ in the original release, but its bandpass includes nebular emission from both \ha\ and the \niiabw\ doublet (Figure~\ref{fig_filter}).

We downloaded the associated High-Level Science Products (HLSPs) from the Hubble Legacy Archive, and used the publicly released drizzled science mosaics for all subsequent analysis (Figure~\ref{fig_field}).
Because these HLSP mosaics were produced prior to Gaia-based astrometric solutions, we updated the ACS astrometry to a common reference frame tied to our new WFC3 imaging. 
We identified a set of isolated point sources detected in both the ACS mosaics and our WFC3/F606W image, measured their pixel coordinates in each image, and solved for an updated WCS using the \texttt{astropy.wcs.utils.fit\_wcs\_from\_points} routine. 
The resulting WCS solution was then applied consistently to all ACS filters. 
Finally, we resampled the astrometry-updated ACS images onto the WFC3 pixel grid to facilitate pixel-by-pixel comparisons and image arithmetic.

To isolate the \nii\ emission, we use the WFC3/F656N \ha\ map as an external constraint. We scale the \ha\ map by the ratio of the WFC3/F656N and ACS/F658N filter widths, and then subtract it from the ACS/F658N image. The remaining emission is treated as the \nii\ map, shown in the second row of Figure~\ref{fig_line_images}.

As with other \emph{HST} instruments, the effective point-spread function (PSF) of ACS and WFC3 depends on the optical PSF, geometric distortion correction, and charge diffusion, and can vary with wavelength and detector position. 
Owing to the absence of sufficiently bright, isolated stars in the region of interest, we do not attempt a full empirical PSF homogenisation across instruments and filters. 
The \nii\ map, constructed from differencing the ACS/F658N and WFC3/F656N images, may therefore contain low-level small-scale systematics associated with PSF mismatch.
This limitation does not affect our filament analysis. As described in Section~\ref{sec_filaments}, filaments are identified exclusively in the H$\alpha$ image, and the \nii\ map is used only to measure the corresponding \nii\ flux within the \ha-defined filament footprints. Moreover, the characteristic transverse widths of the filaments in our analysis ($\sim$7.3 pixels) are substantially larger than the instrumental PSF, so the resulting integrated surface-brightness measurements are insensitive to modest inter-band PSF differences.

\section{Filaments in the Wind} 
\label{sec_filaments}

A central goal of this work is to identify and characterise the filamentary structures embedded within the ionized outflow of M82. In this section, we describe our filament-detection procedure, the construction of a quantitative filament catalogue, and the derivation of key filament properties.

\subsection{Filament detection}
\label{subsec_fil_detect}

A variety of methods have been developed to identify filamentary structure in astronomical data, including topological approaches such as \textsc{DisPerSE} \citep{sousbie11}, multi-scale decomposition methods such as \textsc{getfilaments} \citep{menshchikov13}, Hessian-based algorithms \citep[e.g.][]{schisano14,salji15}, and mask-based skeletonisation methods such as \texttt{FilFinder} \citep{koch15}.
In this paper, we develop a procedure tuned to the high-resolution \emph{HST} \ha\ image and to the goal of separating compact filamentary emission from a physically meaningful diffuse component while remaining robust to noise fluctuations. 
We apply the method to the continuum-subtracted \ha\ map (Section~\ref{subsec_cont_sub}), where the signal-to-noise ratio is highest and the filamentary morphology is most clearly defined. 
The core idea is to enhance small-scale structure by
suppressing large-scale emission through high-pass filtering, and then to link the surviving high-contrast pixels into a coherent filament map using connectivity-based segmentation.
This approach is conceptually similar to \texttt{FilFinder} \citep{koch15}, which was developed and first demonstrated for filamentary structures in molecular-cloud dust-emission maps.
Given the complex emission morphology and the relatively low signal-to-noise ratio in parts of the field, we implement a customised
pipeline tailored to this dataset and to our science goals.

\begin{figure*}
    \centering
    \includegraphics[width=\textwidth]{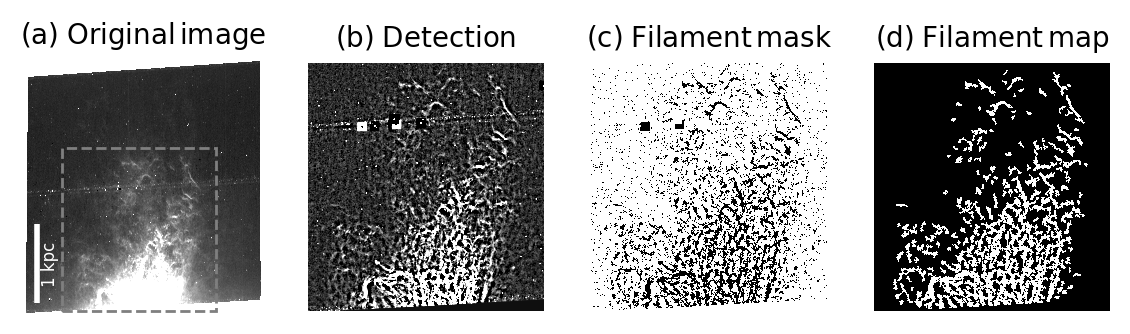}
    \caption{
    The filament-detection procedure described in Section~\ref{subsec_fil_detect}.
    (a) The continuum-subtracted \ha\ image. (b) The structure-enhanced detection image after removal of the large-scale diffuse component. (c) The preliminary filament mask obtained by applying a local S/N threshold and Gaussian smoothing to improve connectivity. (d) The final filament map after linking contiguous pixels, rejecting small spurious regions, and removing obvious artefacts through visual inspection. Panels (b)–(d) show only the dashed rectangular sub-region marked in panel (a).
    }
    \label{fig_fil_detection}
\end{figure*}

We begin by suppressing large-scale diffuse emission in the \ha\ map to enhance filamentary structure. Specifically, we construct a smoothed version of the image by convolving it with a two-dimensional Gaussian kernel of ${\rm FWHM}=25$~arcsec ($\approx$630 pixels at the WFC3/UVIS pixel scale).
This smoothing scale is intentionally much larger than the size of small structures, so that the smoothed map captures only the slowly varying component of the wind emission and any residual large-scale gradients. 
We verified that the resulting filament morphology is stable for Gaussian kernels between 5 and 50~arcsec, while substantially smaller kernels begin to filter emission on scales comparable to individual filaments.
We then generate a high-pass, structure-enhanced image by subtracting the smoothed component from the original, $I_{\rm det}=I_{\mathrm{H}\alpha}-(I_{\mathrm{H}\alpha}\ast G_{\sigma})$, thereby emphasising narrow, high-contrast features while minimising broad diffuse emission.
This step is close in spirit to the unsharp mask approach used in \citet{bolatto24} and \citet{lopez25}.
It produces the detection image shown in Figure~\ref{fig_fil_detection}(b), in which filamentary structures are enhanced against a background.
Unlike \texttt{FilFinder} \citep{koch15}, we do not apply an arctangent intensity-flattening step: in the low-surface-brightness regions of the M82 \ha\ image, compressing the dynamic range before thresholding can make the resulting mask topology sensitive to the flattening and thresholding parameters.

We next identify candidate filament pixels in the detection image by applying a permissive signal-to-noise threshold relative to the local background RMS.
We estimate the noise by adding in quadrature the propagated uncertainty from the drizzled HST products and a local residual RMS measured in $20$-pixel boxes. This initial map is used to identify a preliminary filament mask. We then iterate the same uncertainty estimate after masking the filaments, and use this final noise map for filament selection. Candidate filament pixels are selected from the detection image using a permissive threshold, $I_{\rm det} > 1\,\sigma_{\rm bkg}$, where the median value of the final local threshold is $\sigma_{\rm bkg} \approx 2.5 \times 10^{-16}\,{\rm erg\,s^{-1}\,cm^{-2}\,arcsec^{-2}}$.

We adopt this low threshold intentionally to retain faint filaments and low-contrast bridges that would otherwise be missed by a more conservative cut. The resulting binary mask can contain small holes and discontinuities caused by noise fluctuations or local surface-brightness variations along the filaments. To improve the morphological continuity of the filament network, we therefore convolve the binary mask with a Gaussian kernel of ${\rm FWHM}=10$ pixels.
This step effectively fills small gaps and reconnects nearby high-significance structures, yielding a final connectivity mask that traces coherent filamentary regions; the corresponding mask is shown in Figure~\ref{fig_fil_detection}(c).
This smoothing step sets an effective resolution for separating neighbouring structures: two nearly parallel filaments with separations smaller than the adopted ${\rm FWHM}$ may be merged into a single connected feature. In practice, however, we find that varying the kernel ${\rm FWHM}$ from 5 to 20 pixels does not significantly alter the overall morphology of the resulting filament network.

We then identify connected components in the connectivity mask using 8-connected labelling, which treats diagonal neighbours as connected and is therefore well suited for thin, curvilinear structures. To suppress spurious detections, we remove small regions that are unlikely to represent physical filaments and retain only structures with an area of at least 100 pixels. We further perform a visual inspection of the labelled regions to identify and exclude a small number of obvious artefacts (e.g. residual edge effects, subtraction residuals, or cosmic rays). The output is an integer-labelled region map (Figure~\ref{fig_fil_detection}(d)), providing the basis for subsequent skeletonisation and filament property measurements.

\subsection{Skeletonisation}
\label{subsec_skeleton}

To convert the segmented filament regions into a quantitative catalogue of individual filaments, we derive a one-pixel-wide representation of the filament ``spines'' via morphological skeletonisation. 
This provides a compact description of the filament geometry, enabling us to define filament paths, measure projected lengths, locate junctions, and perform width measurements relative to a well-defined central axis.
The input to this step is the filamentary complex map produced in Section~\ref{subsec_fil_detect}. 

We generate filament spines using the standard morphological thinning operator implemented in \texttt{skimage.morphology.skeletonize}, which reduces each connected region to a one-pixel-wide medial axis while preserving its topology. The output is a binary skeleton map that traces the central paths of the filament networks.

To decompose a network skeleton into individual filament branches, we identify junctions (``knots'') using a local connectivity estimate. We convolve the binary skeleton image with a $3\times3$ kernel of ones and compute 
\begin{equation}
{\rm degree}(\mathbf{x}) = \left[\left({\rm skel}\ast \mathbf{1}_{3\times3}\right)(\mathbf{x}) - {\rm skel}(\mathbf{x})\right],
\end{equation}
such that ${\rm degree}(\mathbf{x})$ counts the number of neighbouring skeleton pixels surrounding $\mathbf{x}$. Pixels with ${\rm degree}\ge 3$ are classified as knots, corresponding to branching or crossing points where multiple filament segments meet, whereas pixels with ${\rm degree}=1$ are identified as endpoints. 
After identifying the knot pixels, we remove them from the network skeletons to break the interconnected skeleton into a set of individual filament branches. 
We do not prune the skeleton to retain only the longest path. The M82 wind forms a connected network of branches and knots, and pruning toward a dominant path could remove substructure that is relevant to the morphology we aim to measure. This choice is also physically motivated. Recent JWST observations of M82 map the cold, dust-bearing cloud by PAH emission, whereas \ha\ preferentially traces ionized skins or mixing layers on cloud surfaces
\citep{lopez26}. The \ha\ image can therefore show sharper interface
substructure than the dust/PAH emission, motivating a branch-based
definition tuned to the \ha\ morphology.
For each filament, we quantify its projected length, defined as the number of pixels in the one-pixel-wide spine.

We estimate a characteristic filament width by combining the filament skeleton with the \emph{exact} Euclidean distance transform (EDT). For each pixel at position $\mathbf{x}$,
we compute its shortest Euclidean distance to the nearest skeleton pixel, $d(\mathbf{x})=\min_{\mathbf{s}\in \mathrm{skeleton}}\|\mathbf{x}-\mathbf{s}\|$, where $\mathbf{s}$ denotes skeleton pixels. 
We then define a single characteristic width for each filament using a flux-weighted mean distance within its footprint,
\begin{equation}
W_{\rm pix} \;\equiv\; 2\,\frac{\sum\limits_{\mathbf{x}\in \mathrm{mask}} F(\mathbf{x})\, d(\mathbf{x})}{\sum\limits_{\mathbf{x}\in \mathrm{mask}} F(\mathbf{x})},
\end{equation}
where $F(\mathbf{x})$ is the continuum-subtracted \ha\ flux at position $\mathbf{x}$. The factor of two converts a characteristic distance from the spine into an effective diameter. By construction, the flux weighting emphasises the bright filament core while down-weighting low-significance boundary pixels, so the resulting width is largely insensitive to small-scale irregularities in the segmentation edge, weak diffuse wings, and local broadening near knots or compact clumps. This resulting width is non-parametric and makes no assumption about the functional form of the transverse surface-brightness profile, providing a robust measure of filament thickness.

The right column of Figure~\ref{fig_line_images} shows four representative regions in the outflow. In each panel, the stretched emission-line map is shown in the background, while the filament skeletons are overlaid as yellow curves. The detection procedure performs well in both the inner and outer outflow, despite the very different levels of diffuse background emission. Although the filaments are detected from the \ha\ map, they closely follow the structures seen in \nii\ and \sii. For \oiii, the emission becomes too faint in the outer wind for a robust comparison, but similar filamentary structures are still visible in the inner wind. This suggests that, where the signal is sufficient, the filamentary morphology is broadly consistent across the different ionized-gas tracers.

\begin{figure}
    \centering
    \includegraphics[width=1.05\columnwidth]{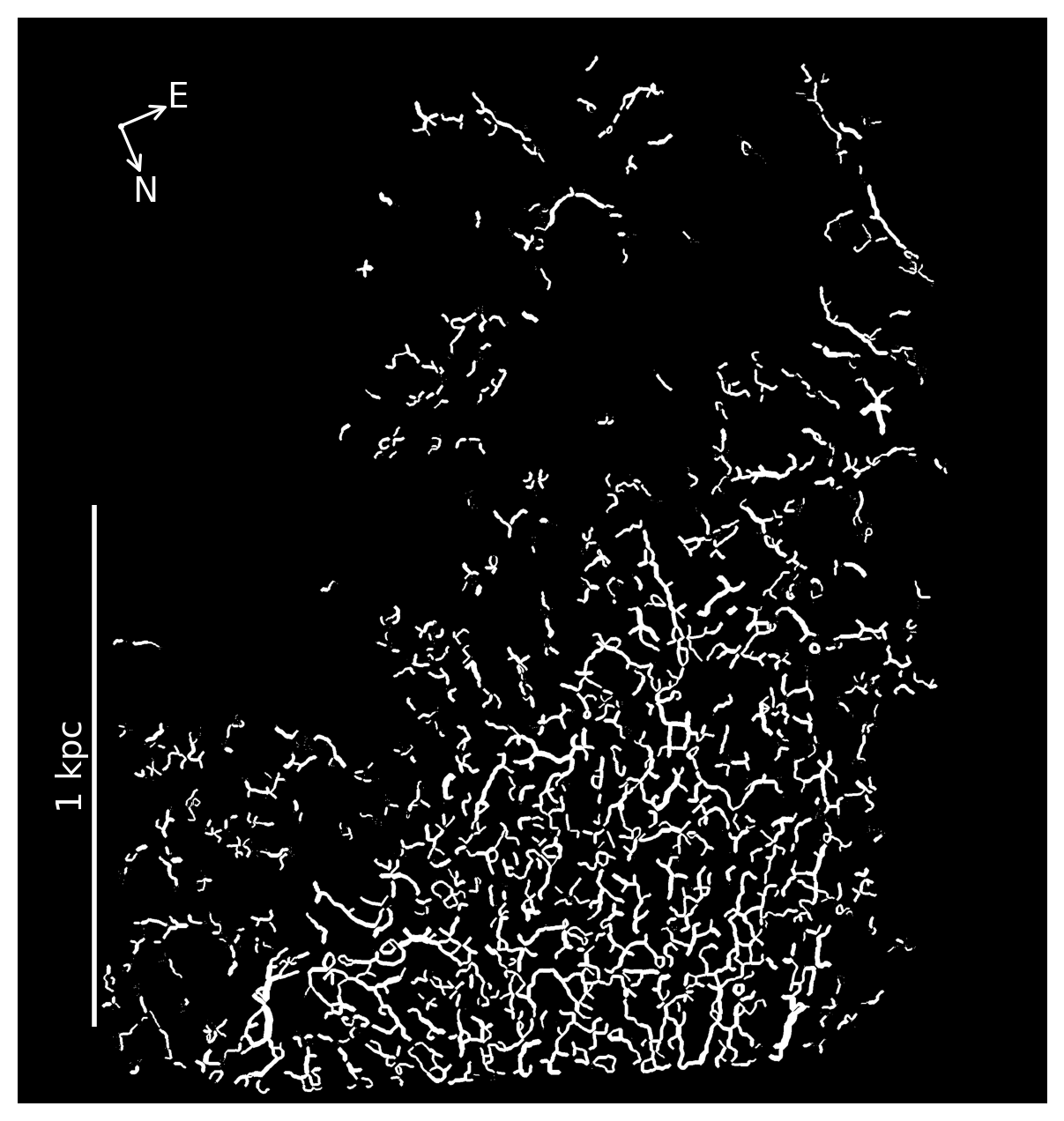}
    \caption{
    The final reconstructed filament skeleton map. For visualisation, each skeleton is rendered with a thickness equal to the measured characteristic filament width (Section~\ref{subsec_skeleton}). 
    }
    \label{fig_skeletons}
\end{figure}

For visualisation, we also reconstruct a filament map by rendering each skeleton segment with a constant thickness set by its measured width. The resulting map reveals a complex, interconnected filament network, as shown in Figure~\ref{fig_skeletons}.
In total, our filament-detection procedure identifies 2526 filaments across the analyzed HST \ha\ field.

The filament distribution in Figure~\ref{fig_skeletons} is not azimuthally symmetric with respect to the galaxy's minor axis (vertical direction on the image), but is instead concentrated along a diagonal toward the northern-eastern side (lower right) of the image. This asymmetry is already visible in the continuum-subtracted \ha\ image before the filament-detection procedure is applied Figure~\ref{fig_original_images}. 
The physical origin is uncertain. It may reflect an asymmetric distribution of recent star formation in the disk, which powers the wind, or an external effect. For example, if M82 has a westward component of motion through the M81 group medium, ram pressure could preferentially displace or compress extraplanar material toward the east.
A related scenario has been proposed by \citet{rao26} to explain the stellar population distribution in the eastern M82 stellar tail.

\subsection{Filament size and morphology}

\begin{figure}
    \centering
    \includegraphics[width=1.0\columnwidth]{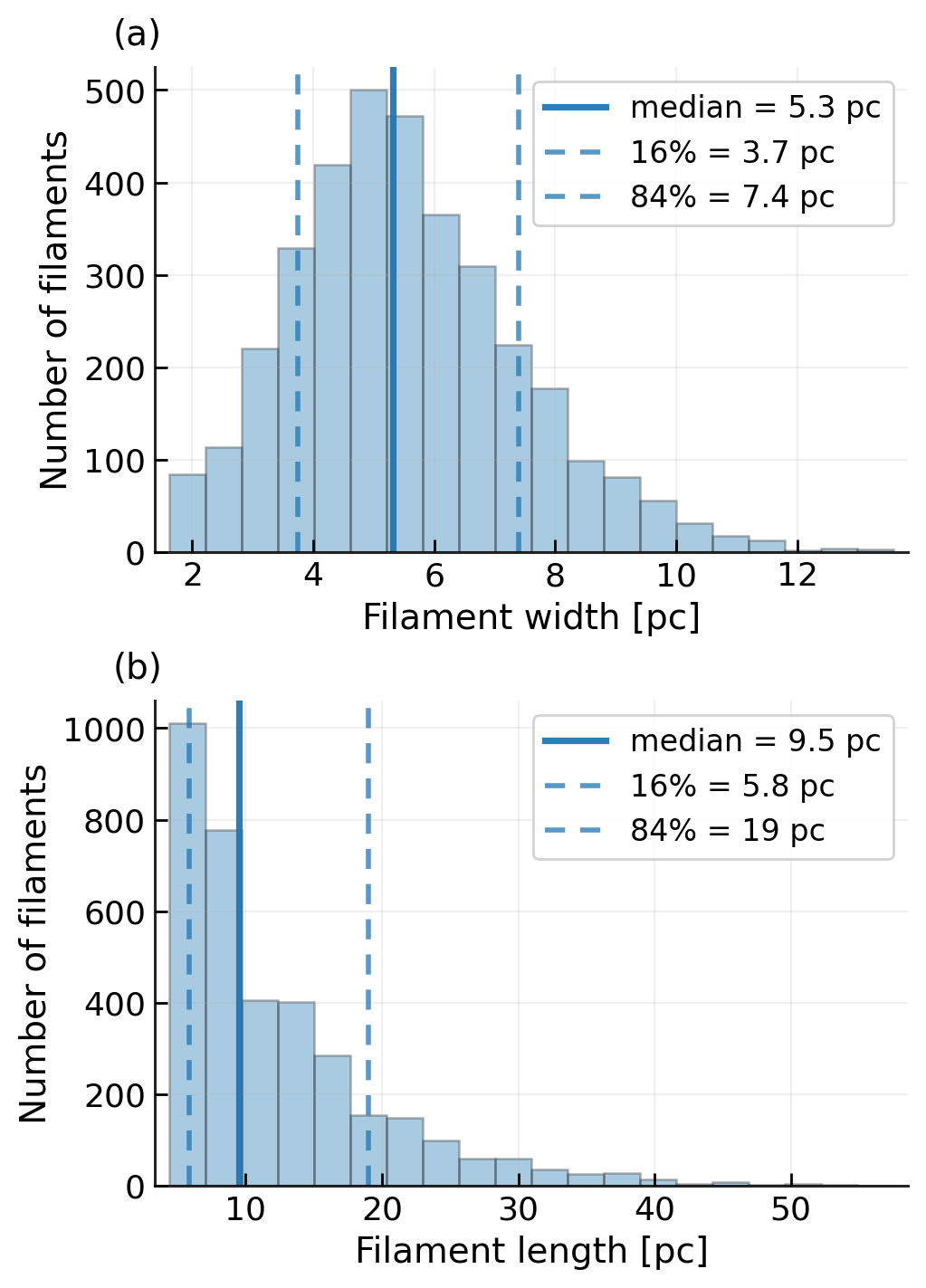}
    \caption{
    Distributions of (a) measured filament widths and (b) projected lengths. Vertical solid lines mark the median values, and dashed lines indicate the 16th and 84th percentiles.  
    The spatial resolution of WFC3/UVIS imaging is about 0.07'' (1.3~pc).
    }
    \label{fig_fil_w_l}
\end{figure}

Figure~\ref{fig_fil_w_l} presents the distributions of filament widths and projected lengths. The width distribution has a median $W \simeq 5.3$~pc (7.3~pixels) and a central 16\% -- 84\% interval of $W\simeq $3.7--7.4~pc. The projected length distribution is skewed towards short length, with a median $L \simeq 9.5$ pc and an interval of $L\simeq$5.8--19~pc, plus a tail extending to $L\gtrsim$40--50~pc. These distributions indicate that the filament network is dominated by numerous compact structures, while a smaller fraction of filaments reaches substantially larger extents.

\begin{figure}
    \centering
    \includegraphics[width=1.0\columnwidth]{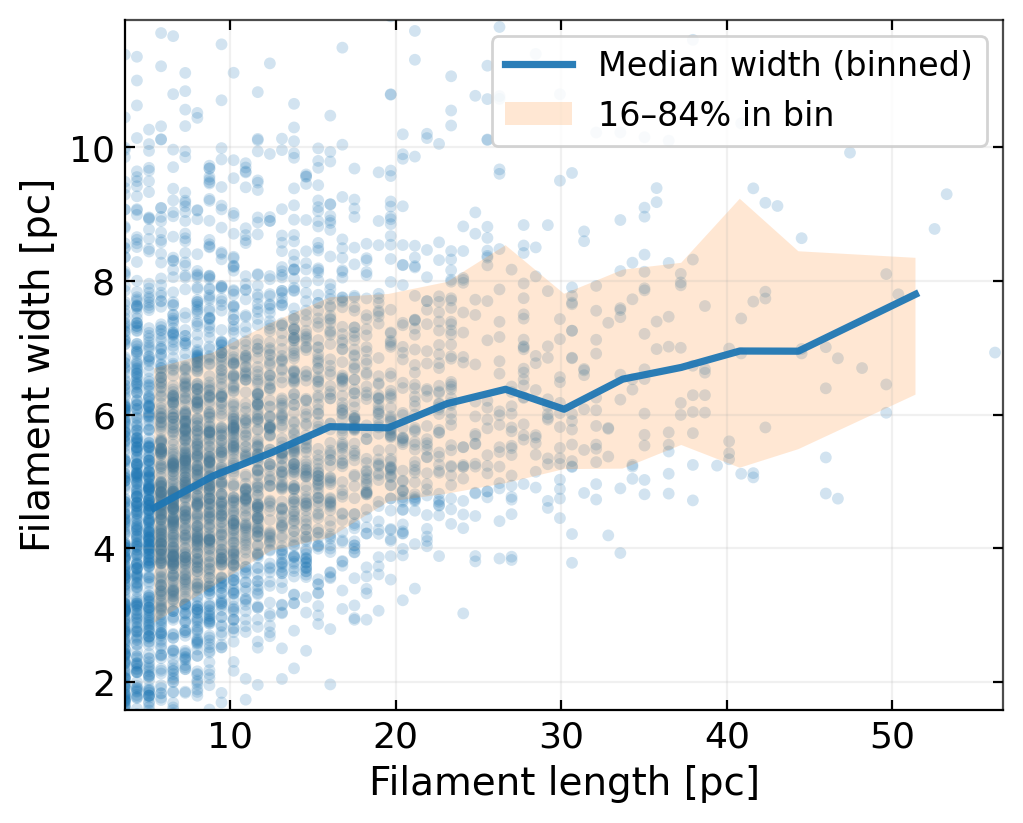}
    \caption{
    Filament width as a function of length. Individual filaments are shown as points, while the solid curve indicates the binned median width; the shaded region illustrates the scatter range within each length bin. The filament width increases mildly with length.
    }
    \label{fig_fil_l_vs_w}
\end{figure}

We examine whether the measured filament properties depend on their location and orientation within the outflow. 
Specifically, we measure each filament’s projected inclination from the overall elongation direction of its skeleton pixels, folded to 0–90 deg relative to the galaxy’s major axis. The inclination distribution is broad, with a median of about 48 deg, and shows no clear alignment with the outflow opening angle.
We find no clear evidence that the filament length, width, or inclination angle evolves systematically with the vertical distance from the base of the outflow, $z$. We also find no significant correlation between filament inclination and either length or width, suggesting that the projected filament morphology is not primarily driven by orientation within the plane of the sky.
We also repeated the analysis using only filaments longer than 20 pc, which removes many short sub-branches from fragmented skeletons, and still find no clear correlation between filament properties and distance from the disk.
The clearest, albeit still weak, trend is between filament width and length. Figure~\ref{fig_fil_l_vs_w} shows that the distribution has substantial scatter at all lengths, reflecting the diversity of filament morphologies. Nevertheless, the binned median width increases mildly with filament length. Longer filaments are, on average, slightly broader, but length remains a poor predictor of width for individual structures.

When interpreting these morphological measurements, several caveats should be kept in mind. First, the filament width and length are defined using different methods. The length is a projected quantity measured from the filament graph. If a long branch contains several smaller sub-branches, the structure can be broken at the connecting knots and recorded as multiple shorter filaments. The width, however, is measured from the flux-weighted transverse size and therefore traces a different aspect of the morphology. Second, our filament identification is inherently two-dimensional. Some structures that appear connected in projection may not be physically connected in three dimensions, while filaments inclined away from the plane of the sky will have true lengths larger than their projected values. These effects can naturally produce cases in which the measured width is comparable to, or even larger than, the measured length (Figure~\ref{fig_fil_l_vs_w}). 
In this sense, the measured filament width is likely to be a more robust and directly interpretable quantity than the projected filament length.

\begin{figure}
    \centering
    \includegraphics[width=1.0\columnwidth]{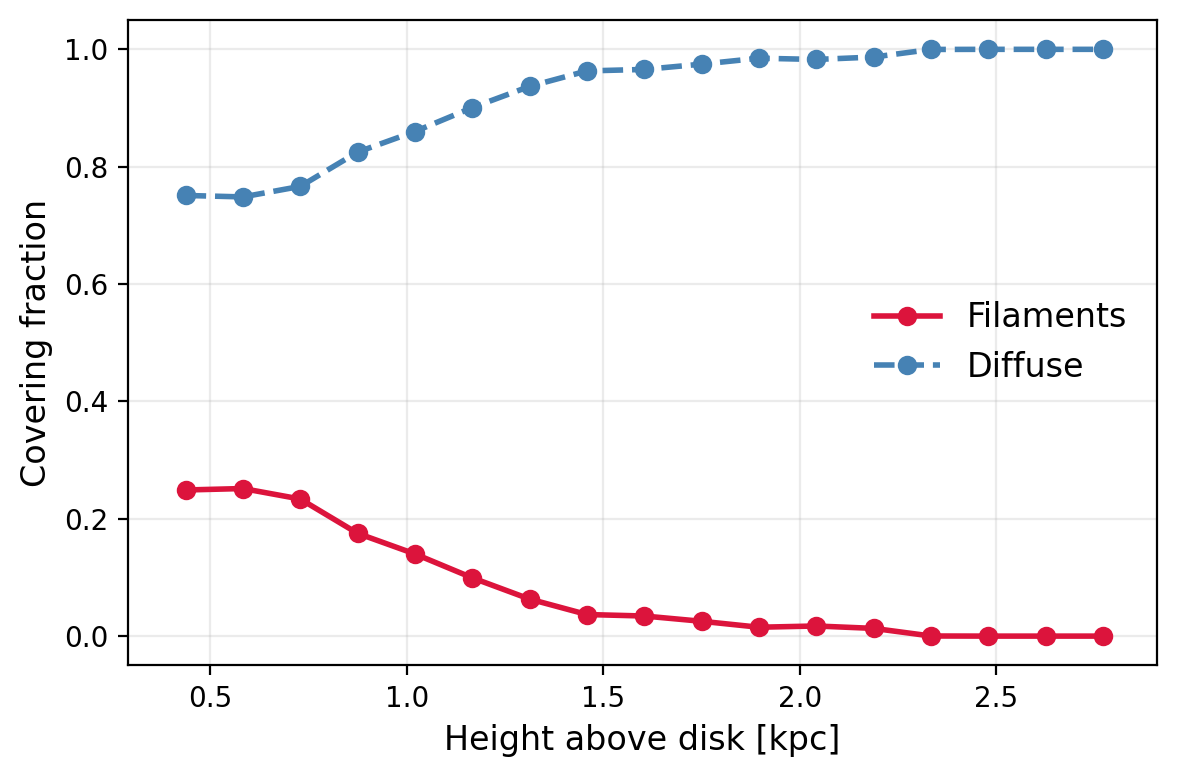}
    \caption{
    Covering fraction measured in successive conical radial bins as a function of height above the disk, for the filamentary component (red) and the complementary diffuse emission (blue). The filamentary covering fraction declines rapidly with height, showing that the filament network is concentrated in the inner wind, while the outer wind is increasingly dominated in projected area by diffuse emission.
    }
    \label{fig_fil_cf}
\end{figure}

In addition to the filament-by-filament morphology, we quantify how the areal covering fraction of filamentary versus diffuse emission varies with height above the disk (Figure~\ref{fig_fil_cf}). 
We first restrict the analysis to an outflow ``cone'' centred on the minor axis, adopting an opening angle of approximately $70^\circ$ as measured by \citet{xu23}. 
Within successive conical radial bins, we compute the fraction of unmasked pixels classified as filaments, $f_{\rm cov}^{\rm fil}(z) \equiv \frac{A_{\rm fil}(z)}{A_{\rm tot}(z)} $,
where $A_{\rm fil}(z)$ is the number of pixels in the filament mask and $A_{\rm tot}(z)$ is the total number of pixels within the conical bin at that height. 
The complementary diffuse covering fraction is defined as $f_{\rm cov}^{\rm diff}(z) \equiv 1 - f_{\rm cov}^{\rm fil}(z)$,
i.e.\ the fraction of pixels assigned to the non-filament (diffuse) component within the same region.

There is a pronounced transition with height: near the base of the WFC3 field ($z \simeq 0.4$--0.6 kpc), filaments cover $\sim$25\% of the area, but this fraction declines steadily with increasing $z$, dropping below $\sim$3\% by $z \sim$1.5 kpc and approaching zero beyond $z \gtrsim 2$ kpc. 
This behaviour suggests that the small-scale filamentary network is concentrated in the inner wind, whereas the outer regions are morphologically dominated by diffuse emission. 
We caution, however, that we cannot rule out the presence of fainter \ha\ filaments at larger heights that fall below our detection threshold.
We also examined the distribution of skeleton knots, which trace branching or intersection points in the filament network, and find that their number density decreases with distance in a manner broadly similar to the filament covering fraction. This indicates that such points are concentrated where the filament network has the highest areal covering fraction.

\subsection{Filament emission}
\label{subsec_sb}

\begin{figure}
    \centering
    \includegraphics[width=1.0\columnwidth]{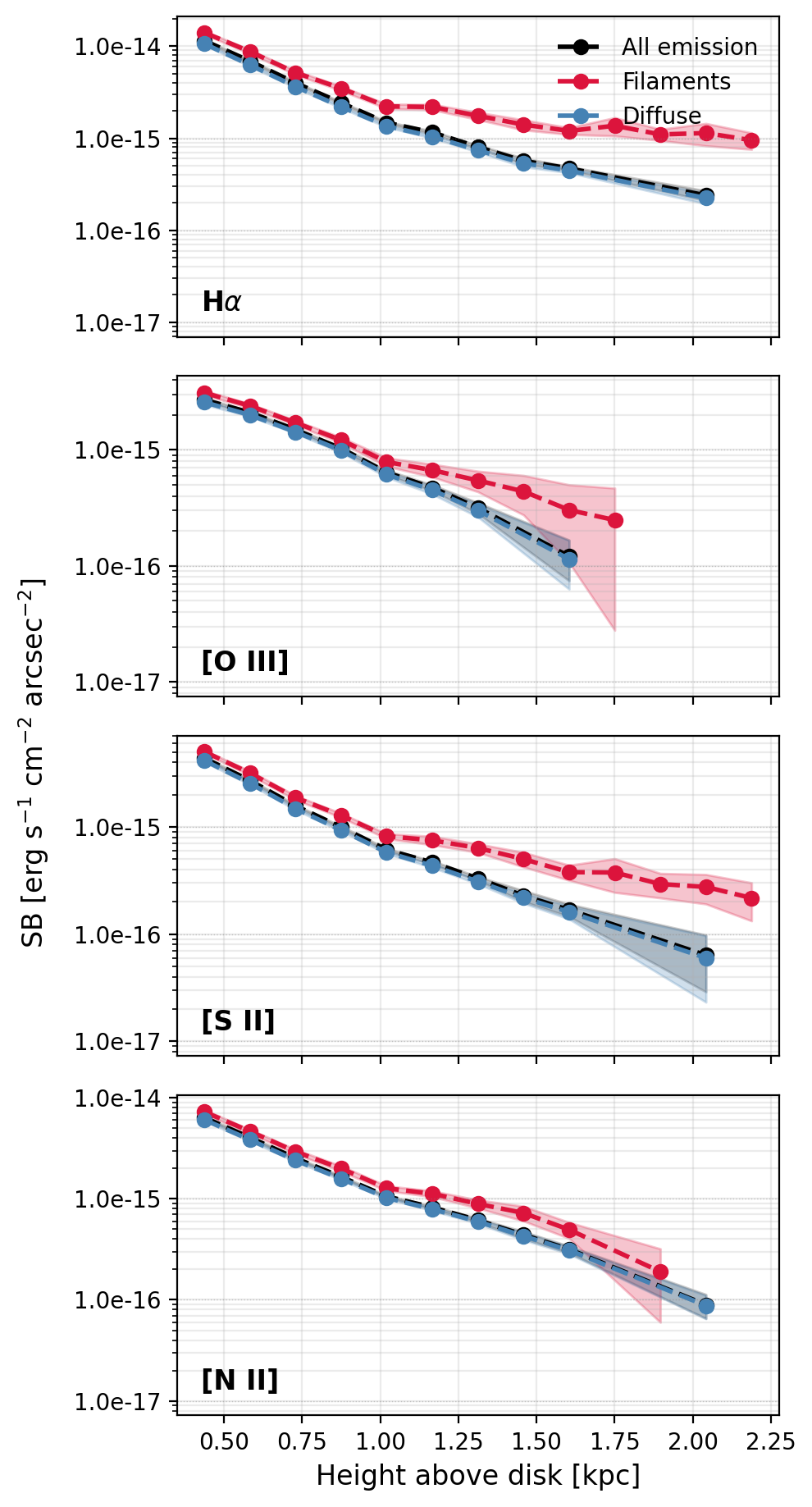}
    \caption{Median surface brightness profiles measured within the outflow cone for \ha, \oiii, \sii, and \nii, shown for the total emission as well as the filamentary and diffuse components. Only bins with ${\rm S/N}>2$ are plotted. Shaded bands indicate the $1\sigma$ uncertainties. Note that the contrast between the surface brightness of the filaments compared to the diffuse emission increases with distance.}
    \label{fig_sb_profiles}
\end{figure}

To assess the radiative output of the warm ionized gas, we measure vertical surface-brightness profiles for the principal optical lines within the adopted outflow cone from the narrow band images (Figure~\ref{fig_line_images}).
Figure~\ref{fig_sb_profiles} shows the median surface-brightness profiles measured in successive bins within the adopted outflow cone, plotted as a function of height above the disk, for the total emission (filament$+$diffuse) and for the filamentary and diffuse components separately.
Both filamentary and diffuse emission fade with height, with line-dependent differences in the radial extent (in particular, \oiii\ becomes noise-limited at larger heights).

\begin{figure}
    \centering
    \includegraphics[width=1.0\columnwidth]{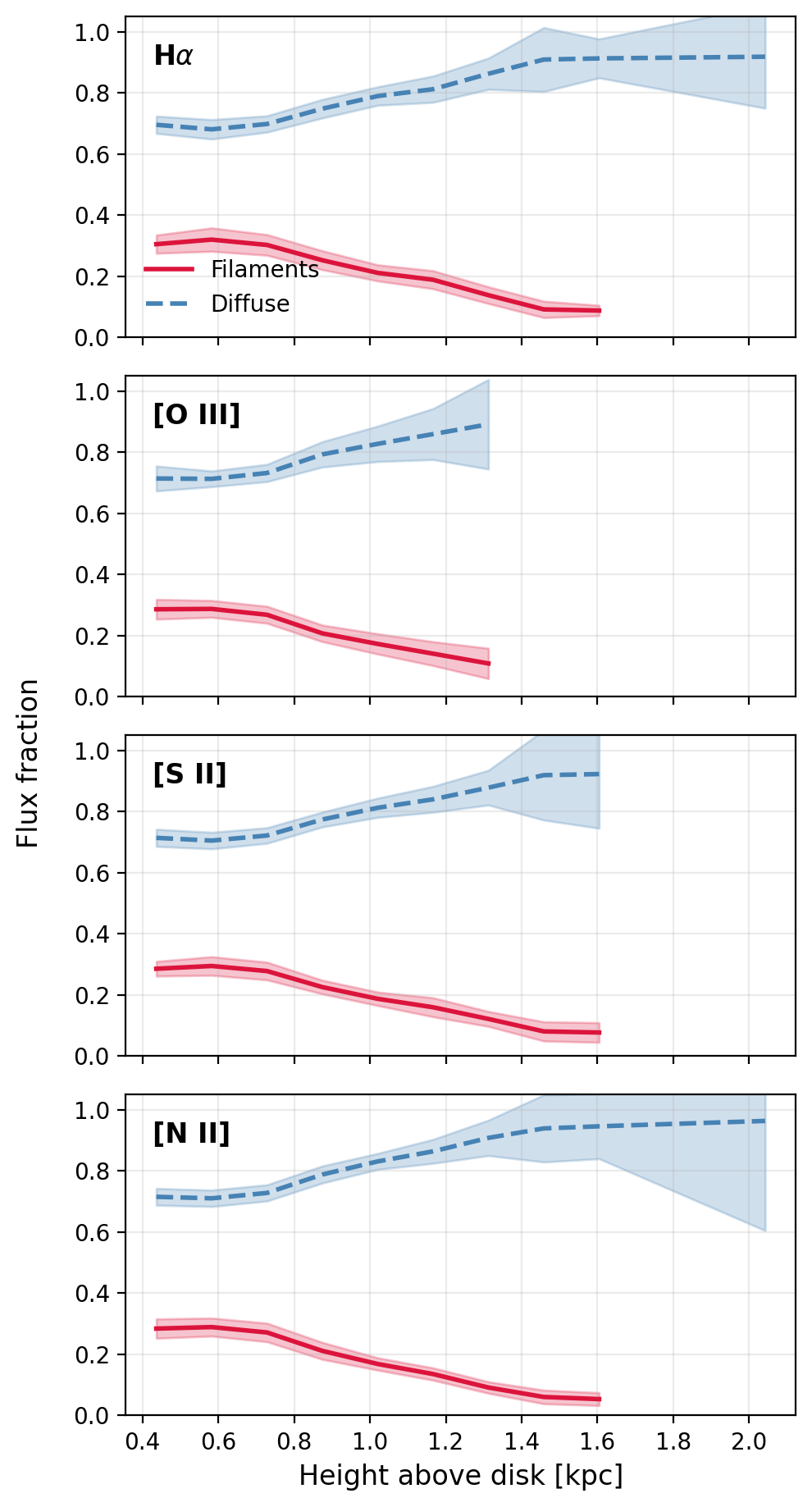}
    \caption{
    Flux fractions of the filamentary and diffuse components measured in successive conical radial bins as a function of height above the disk. Only bins with ${\rm S/N}>2$ are shown. Shaded bands indicate the $1\sigma$ uncertainties. Filaments contribute only a minority of the total line flux, and their contribution declines with height.}
    \label{fig_f_frac_profiles}
\end{figure}

Although the diffuse component occupies most of the projected area (Figure~\ref{fig_fil_cf}), filaments remain locally bright and show high surface brightness.
However, filaments do not contribute substantially to the total flux. This is quantified in Figure~\ref{fig_f_frac_profiles}, which shows the fraction of the line flux arising from the filament versus the complementary diffuse area as a function of height. Near the base of the cone ($z\sim0.4$--$0.7$ kpc), filaments contribute $\sim$25--35\% of the flux, with the remaining $\sim$65--75\% emitted by the diffuse component.
With increasing height, the filament flux fraction declines steadily in all lines, reaching $\lesssim$10\% by $z\gtrsim1.4$--$1.6$ kpc (and remaining low at larger heights in \ha, where the profile extends furthest). Correspondingly, the diffuse component dominates the radiative output at large heights. However, Figure~\ref{fig_sb_profiles} shows that the ratio of the surface brightnesses of the filaments to those in the diffuse emission increases with increasing height, especially beyond 1 kpc.

Taken together, Figures~\ref{fig_sb_profiles} and \ref{fig_f_frac_profiles} imply that, while filaments trace visually striking, high-contrast structures, the bulk of the optical line luminosity, and hence a substantial fraction of the optical line radiative output of the warm ionized phase, is carried by the diffuse component, especially beyond the inner wind. A natural interpretation is that \ha\ and the forbidden-line emission largely arise in mixing/ionization layers that become increasingly diffuse with height, whereas coherent filaments represent a decreasing fraction of the emitting interfaces as structures expand, fragment, or are disrupted during propagation through the wind.

\subsection{Excitation mechanisms}
\label{subsec_excitation}

\begin{figure*}
    \centering
    \includegraphics[width=2.0\columnwidth]{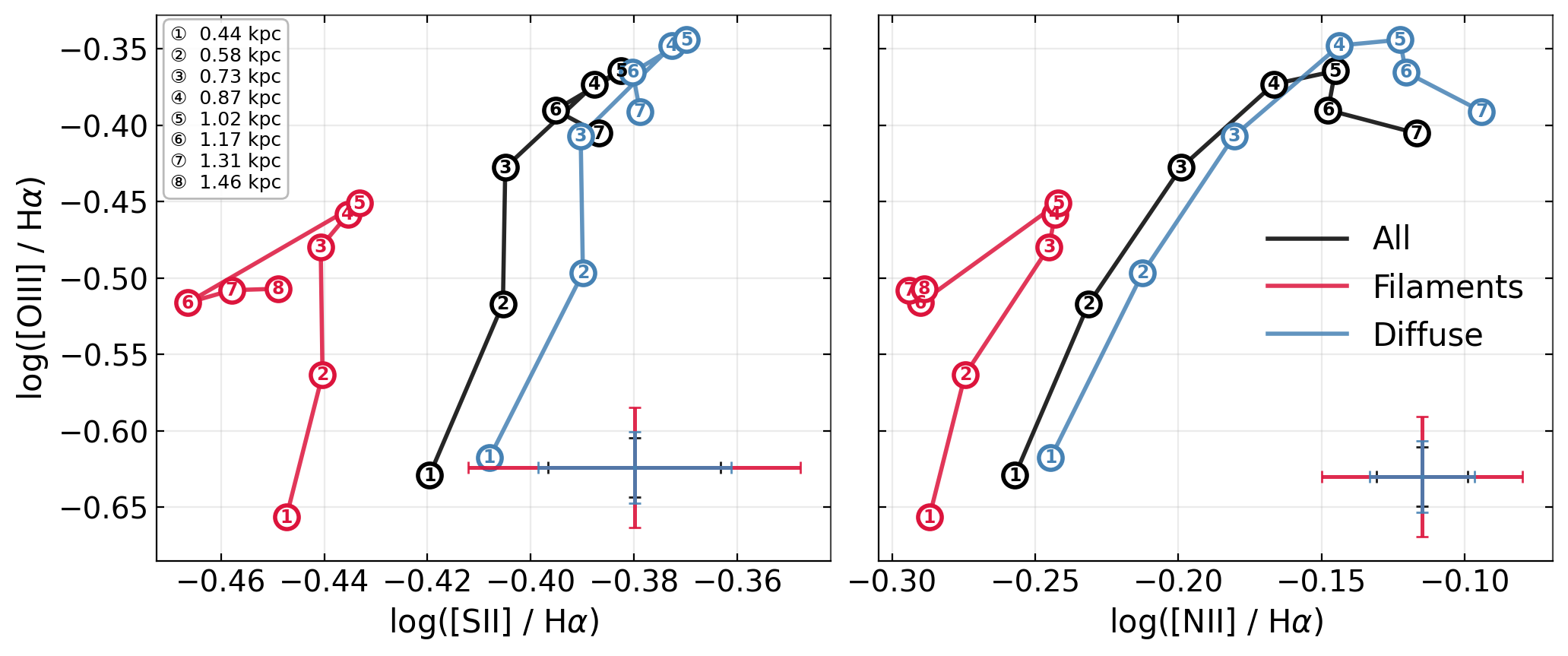}
    \caption{    
    Optical diagnostic line-ratio diagrams for the M82 outflow cone, showing \oiii /\ha\ versus \sii/\ha\ (left) and \nii/\ha\ (right). 
    Points show measurements in successive conical radial bins and are colour-coded by component (filamentary, diffuse, and total); labels indicate the representative height of each bin.
    Despite the uncertainties, the filamentary and diffuse components occupy distinct regions of line-ratio space. All components also show a systematic evolution with vertical distance, tracing changes in the excitation state along the outflow.
    }

    \label{fig_diagnostics}
\end{figure*}

To constrain the dominant excitation mechanisms of the warm ionized gas, we examine line ratios as a function of height above the disk, separating the emission into filamentary and diffuse components (Figure~\ref{fig_diagnostics}). 
We consider the diagnostic planes \oiiiw /\ha\ versus \siiabw /\ha\ and \oiiiw /\ha\ versus \niibw /\ha. 
Points are colour-coded by component (filaments, diffuse, and total) and annotated by the representative height of each conical radial bin, enabling us to trace systematic changes in excitation along the outflow.
The filamentary and diffuse components show  offsets in both diagnostic diagrams, despite the large uncertainties. The diffuse emission shows systematically higher \sii/\ha\ and \nii/\ha, whereas the filamentary component lies at lower \sii/\ha\ and \nii/\ha. The total (``all'') line ratios fall between these two components but are generally closer to the diffuse component, consistent with the diffuse component dominating the total line flux at most heights (Section~\ref{subsec_sb}).

A coherent vertical evolution is also apparent. Up to $\sim1$~kpc, \oiii /\ha\ increases with height for all components, indicating progressively higher excitation conditions in the inner wind. Beyond $\sim1$~kpc, the filamentary and diffuse components diverge more clearly. The filamentary emission shows relatively modest changes in \oiii /\ha\ and shows offset to lower \sii /\ha\ and \nii /\ha, whereas the diffuse component drifts toward higher \sii /\ha\ and \nii /\ha\ at roughly unchanged \oiii /\ha.

\begin{figure*}
    \centering
    \includegraphics[width=2.0\columnwidth]{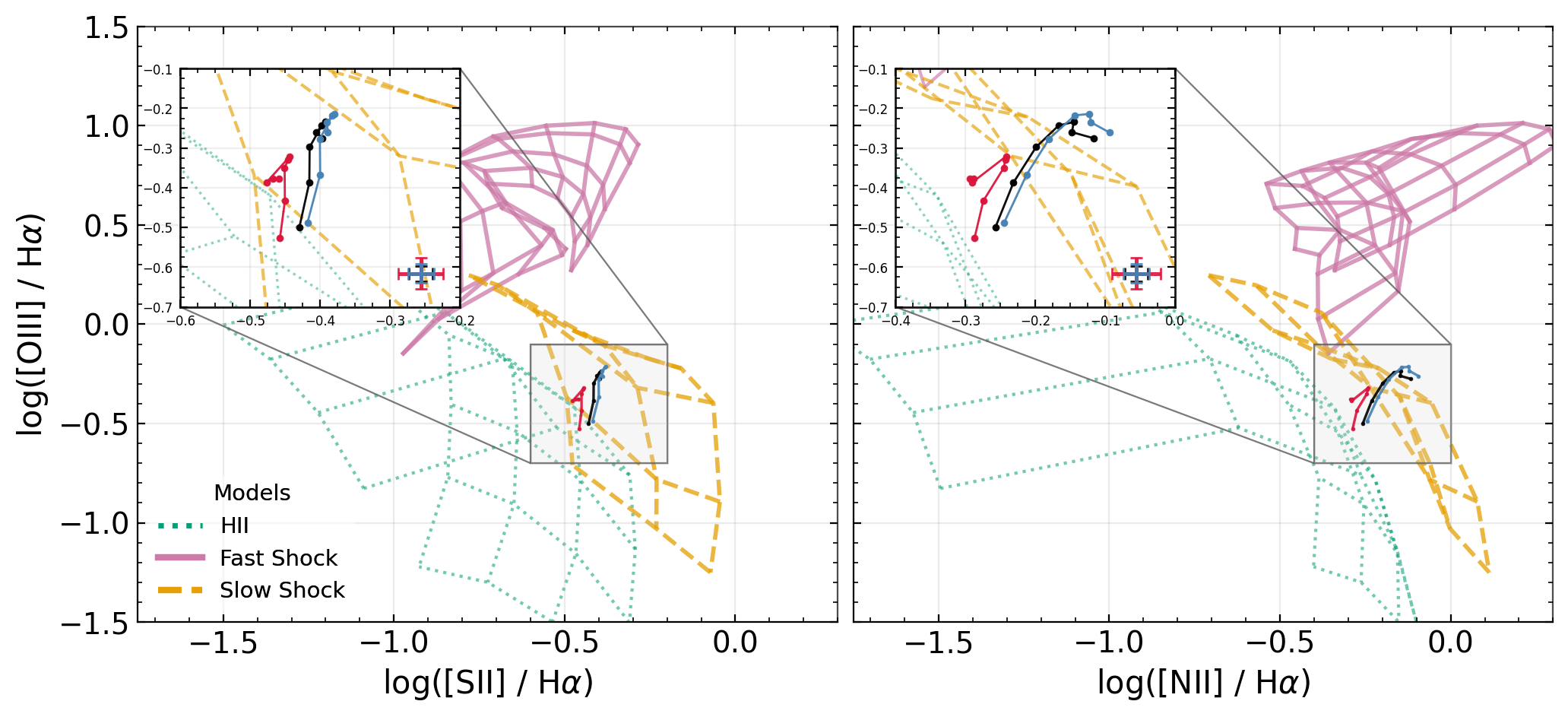}
    \caption{Comparison of observed line ratios with theoretical excitation models in the \oiii/\ha\ versus \sii/\ha\ (left) and \nii/\ha\ (right) diagnostic diagrams. Theoretical model grids are shown for the H\,\textsc{ii}-region photoionization (green), slow shocks (orange), and fast shocks plus precursor (purple). Data points and colour coding follow Figure~\ref{fig_diagnostics}, with the observed \oiii /\ha\ ratios corrected for dust extinction (0.13~dex) using the Balmer decrement from \citet{heckman90}. Insets show a zoomed view of the region occupied by the observations. The data lie in a transition region between the photoionization and slow-shock grids, with the diffuse component migrating progressively towards the shock models at larger heights above the disk.}
    \label{fig_bpt_models}
\end{figure*}

We overlay our line-ratio measurements on grids of photoionization and shock models in Figure~\ref{fig_bpt_models}.
The H\,\textsc{ii} photoionization model is adopted from \citet{li24}, and shock models are adopted from \citet{sutherland17}.
They are computed with \textsc{mappings}~v5.2 \citep{sutherland18}, using the abundance set of
\citet{nicholls17} and the dust depletion factors of \citet{jenkins09}.
Dust destruction is also included in shock models.
A gas pressure of $\log(P/k)\approx 6.2$ is adopted for all three grids, motivated by direct pressure measurements in the M82 wind from \citet{xu23} across the radial range probed by our \textit{HST} imaging.  
Three model grids are included. 
(i) \textit{H\,\textsc{ii}-region photoionization} (green) spans a range of ionization parameters
$-3.5 \leq \log U \leq -2.8$ and metallicities
$7.8 \leq 12 + \log(\mathrm{O/H}) \leq 9.3$, 
with stellar ionizing field derived from Flexible Stellar Population Synthesis \citep[FSPS,][]{conroy09,conroy10} using the Stromlo Stellar Tracks \citep{grasha21} with a Chabrier IMF (Chabrier 2003) and a continuous star-formation history at 5~Myr.
(ii) \textit{Slow shocks} (orange;
$160 \lesssim v_\mathrm{s} \lesssim 200\ \mathrm{km\,s^{-1}}$) represent purely post-shock cooling emission.
(iii) \textit{Fast shocks} (purple; $260 \lesssim v_\mathrm{s} \lesssim 1100\ \mathrm{km\,s^{-1}}$) include significant emission from the photoionized precursor gas \citep{sutherland17}, 
with the post-shock and precursor components each contributing 50\% of the total H$\beta$ luminosity.

As shown in Figure~\ref{fig_bpt_models}, the observed line ratios occupy a relatively small region in both diagnostic planes, confined to a transition zone between the photoionization and shock grids.
Progressing from lower to higher heights above the disk, the data points migrate from the H\,\textsc{ii} region towards the region bounded by the slow-shock grid. 
The M82 wind is therefore neither purely photoionized nor dominated by energetic shocks, but instead lies in an intermediate excitation state.

The filamentary component remains closer to the H\,\textsc{ii}-region photoionization grid. This suggests that the compact, high-surface-brightness strands are plausibly dominated by cloud surfaces illuminated by ionizing photons leaking from the central starburst along the outflow cone. In this picture, \ha\ traces recombination in photoionized gas, while \sii\ preferentially arises near the outer, partially ionized zones and \nii\ traces an intermediate ionization zone \citep{kewley19}. 
The spatial correspondence among \ha, \sii\ and \nii\ (Figure~\ref{fig_line_images}) is therefore consistent with these lines arising from related ionized interfaces, although the present imaging cannot resolve the detailed ionization stratification across individual filaments.

The diffuse component, by contrast, traces more pervasively extended gas that may be more directly exposed to the surrounding hot wind fluid or to a harder ionizing radiation field. 
It shows systematically enhanced \oiii/\ha, \sii/\ha, and \nii/\ha\ relative to the filamentary component, indicating that the diffuse gas occupies a distinct excitation regime rather than simply representing a lower-surface-brightness extension of the filaments. At large heights above the disk, these elevated line ratios may reflect additional non-photoionization heating, such as shocks or turbulent mixing layers, driven by the interaction between the warm gas and the hot outflow \citep{thompson24}. The present narrow-band imaging does not provide the kinematic information needed to isolate shock-excited gas, but the migration of the diffuse emission towards the shock-model grids with increasing height (Figure~\ref{fig_bpt_models}) suggests that such additional heating may become increasingly important at larger distances from the disk.

As discussed in Section~\ref{subsec_skeleton} and shown in Figure~\ref{fig_skeletons}, the \ha\ filaments are preferentially concentrated along a northeastern diagonal. The origin of this asymmetry is uncertain. It may reflect an uneven distribution or history of star formation in the disk, or an environmental effect associated with the relative motion between M82 and the surrounding gaseous medium. 
We do not find clear evidence of the east--west asymmetry in the optical line ratios within the S/N limits of the fainter lines.

\subsection{Energy budget}
\label{subsec_energy_budget}

As shown in Section~\ref{subsec_excitation}, the optical line ratios of the M82 outflow lie in a transition region between the photoionization and shock excitation grids. This indicates that shocks or other non-photoionization processes may contribute to the excitation, especially in the diffuse component at large heights. However, the integrated \ha\ luminosity is dominated by the inner wind: integrating the \ha\ surface-brightness profile in Figure~\ref{fig_sb_profiles} within an outflow cone shows that approximately 50\% of the total \ha\ flux arises within the inner $\sim 0.7$~kpc. Since this region lies closest to the central starburst and remains broadly consistent with photoionization, it is important to test whether Lyman-continuum photons escaping along the outflow cones can provide the required ionizing budget.

We therefore estimate the ionizing photon rate required to power the observed \ha\ emission. The \ha\ surface brightness profile (Figure~\ref{fig_sb_profiles}) is corrected for dust extinction using the \ha/\hb\ ratio from \citet{heckman90}. 
For each conical bin, we convert the extinction-corrected surface brightness profile into luminosity by multiplying the corresponding projected area,  
adopting a bin size of 200 pc along the outflow axis
and a lateral width set by the observed outflow opening angle. We then integrate over $z=400$--2200~pc where the \ha\ surface brightness profile is detected (Figure~\ref{fig_sb_profiles}) and multiply by a factor of two to account for the north-western half of the bipolar outflow.

For our adopted distance to M82, this procedure gives a total outflow \ha\ luminosity of $L_{H\alpha}=1.05\times10^{41}$~erg~s$^{-1}$. 
Under Case~B recombination, this corresponds to a required ionizing photon rate of $Q_{\rm outflow}=\frac{L_{H\alpha}}{1.36\times 10^{-12} erg}=7.7\times10^{52}$~s$^{-1}$ \citep[e.g.][]{osterbrock06}. 
For the total star-formation rate of M82, ${\rm SFR}=8.5~M_{\odot}~{\rm yr}^{-1}$ \citep[e.g.][]{heckman17}, the expected production rate of Lyman-continuum photons is $Q_{\rm tot}=1.4\times 10^{53} \, SFR=1.2\times10^{54}$~s$^{-1}$\citep{kennicutt12}. 
The combined solid angle of the two outflow bicones is approximately $0.2\times4\pi$~sr. 
If the ionizing photons are emitted isotropically by the central starburst, the photon rate initially directed into the two bicones is therefore $Q_{\rm bicones}=2.4\times10^{53}$~s$^{-1}$. 
The observed outflow requires $Q_{\rm outflow}\simeq0.32\,Q_{\rm bicones}$. 
For this ionizing photon rate distributed over the bicone solid angle, the ionization parameter is
\(U = Q_{\rm outflow}/(\Omega_{\rm bicone} n_e r^2 c)
\sim 10^{-3}(n_e/100~{\rm cm}^{-3})^{-1}(r/{\rm kpc})^{-2}\),
where \(\Omega_{\rm bicone}=0.2\times4\pi\).
This value for $U$ agrees with the grid of photoionization models in Figure~\ref{fig_bpt_models}.

This comparison shows that photoionization by the central starburst is energetically feasible, provided that at least $\sim32\%$ of the ionizing photons initially emitted into the bicones escape the inner starburst region and are absorbed by gas between $z=400$ and 2200~pc. Equivalently, the outflowing gas must have a relatively clear line of sight to the starburst radiation field. The remaining $\sim68\%$ of the ionizing photons emitted into the bicones could either be absorbed by gas and dust interior to 400~pc, or escape to radii larger than 2.2~kpc.\footnote{For this bi-cone geometry, $\sim80\%$ of the total ionizing photons produced by the starburst are not directed into the bicones and will instead intercept the starburst disk gas, photoionizing it and powering infrared dust emission.} Thus, although the line-ratio diagnostics suggest that shocks or other additional heating processes may contribute locally, the global \ha\ luminosity does not require shocks as the dominant power source.

\section{Discussion} 


A long-standing unknown in galactic-wind physics is the characteristic morphology and size of the small structures entrained in the outflow, and how these structures evolve as they propagate away from the disk.
In this work, we use deep, high-resolution \emph{HST} narrow-band imaging to directly resolve the small-scale structure in the southern (approaching) outflow of M82. 
The warm ionized emission is highly filamentary, forming a complex interconnected network. 
We measure typical projected filament dimensions of length $\simeq 9.5$~pc and width $\simeq 5.3$~pc, with substantial intrinsic scatter and only a mild tendency for longer filaments to be slightly broader. 
Because filament lengths can be affected by projection and by how individual knots and branches are segmented within a connected network, we emphasise that the width measurements are more physically meaningful: they are defined non-parametrically from the distance transform and are therefore insensitive to subjective choices in skeleton pruning and branch assignment.

\subsection{Comparisons to the literature}

In the literature, different tracers and inference methods probe different phases and different sub-components of the multiphase complexes. Using Subaru/FOCAS \ha\ imaging combined with \sii\ density constraints, \citet{xu23} inferred a very small characteristic radius for the warm ionized clouds ($R_{\rm cl}\simeq 0.07$--$0.9$\,pc) by estimating a volume filling factor within a conical outflow and adopting a line-of-sight covering factor from low-redshift absorption systems. 
These radii are well below the spatial resolution of HST imaging and are $\gtrsim$1 order of magnitude smaller than the characteristic widths that we measure for \ha\ filaments. A likely explanation is that the \citet{xu23} constraint is not measuring the transverse size of the morphological filaments seen in emission, but rather the effective size of the densest emitting substructures that dominate $n_{\rm e}^2$-weighted diagnostics and are required by their filling-factor/covering-factor model. In this sense, their inferred sub-pc ``clouds'' could correspond to unresolved clumps or ionized skins embedded within (or decorating the surfaces of) larger, parsec-to-tens-of-parsec filamentary complexes. 

\citet{lopez25} report prominent arc-like structures identified in the archival ACS/F658N images \citep{mutchler07}, with characteristic sizes of $\sim$24--110~pc. They interpret these features as bow shocks or shell-like interfaces produced as cool material is overrun by the hot wind. Such large arcs need not correspond to single coherent clouds; rather, they may arise from ensembles of smaller clouds and filaments whose collective interaction with the wind produces a contiguous shocked/ionized surface that appears as an arc in projection.

\citet{fisher25} used JWST NIRCam 3.3\,$\mu$m PAH imaging near the base of the outflow and identified coherent ``plumes'' with widths of order $\sim$30--40\,pc and heights larger than $\sim$300--400\,pc, within which they further decomposed smaller clouds with widths of $\sim$5--18\,pc. 
The latter are in reasonable agreement with our filament widths, although \ha\ and PAH emission need not trace identical characteristic scales.
At the same time, the presence of larger plume structures supports a hierarchical picture in which many parsec-scale clouds/filaments are bundled into larger launch channels.
A comprehensive filament-detection and morphological comparison between the PAH-emitting and warm ionized gas will be carried out in future work, in order to determine whether these tracers follow the same underlying cloud hierarchy or instead trace different layers of the multiphase outflow.

This multiscale picture is qualitatively consistent with recent high-resolution simulations, in which cool/warm gas in galactic winds is continuously reshaped by its interaction with the hot wind, producing fragmented structures and dense cloudlets \citep[e.g.][]{schneider17}. Global M82-like simulations further show that the survival, acceleration, and covering fraction of this cool/warm phase depend on both multiphase mixing and the geometry of stellar feedback \citep{schneider20}.
Our measured \ha\ filament lengths and widths most plausibly trace the transverse scales of ionized interfaces, rather than the smallest cloudlets that may dominate density-sensitive diagnostics. The weak increase of width with length, together with the large scatter, suggests a hierarchical and evolving structure rather than a population with a single characteristic scale. This is consistent with models in which superbubble breakout fragments the multiphase ISM and seeds the wind with cool structures spanning a broad range of sizes \citep{tan24}. In such models, the largest structures are set by the characteristic ISM scale, while the low-size end is regulated by the competition between hydrodynamic disruption and radiative cooling in the mixed gas.
The observed decline in filament covering fraction with height may therefore reflect genuine structural evolution (e.g.\ dispersal and shredding into lower-contrast substructure through turbulent mixing), although sensitivity and contrast limitations at low surface brightness likely contribute to the trend.

The absence of a significant trend in filament width, length, or inclination angle with radial distance $z$ is noteworthy in the context of hydrodynamic instability and cloud survival. 
Classical cloud-crushing arguments predict that dense structures embedded in a fast, hot wind are subject to Kelvin--Helmholtz (KH) instabilities at the cloud--wind interface, with a characteristic timescale
$t_{\rm cc} \sim (r_{\rm cl}/v_{\rm wind})\sqrt{\chi}$, where $r_{\rm cl}$ is the characteristic cloud radius, $v_{\rm wind}$ is the hot-wind velocity, and $\chi=\rho_{\rm cl}/\rho_{\rm wind}$ is the density contrast \citep{klein94}. 
For M82-like conditions, the \ha-emitting clumps have densities of
\(n_{\rm cl}\sim100~{\rm cm^{-3}}\) \citep{yoshida19,xu23}, while the volume-filling hot wind has a characteristic density of
\(n_{\rm wind}\sim0.1~{\rm cm^{-3}}\) and velocity of
\(v_{\rm wind}\sim1500~{\rm km~s^{-1}}\) \citep{strickland09}. These values imply a density contrast of
\(\chi=n_{\rm cl}/n_{\rm wind}\sim10^{3}\); for a characteristic cloud radius of \(\sim5\)~pc, the resulting cloud-crushing time is only
\(t_{\rm cc}\sim0.1\)~Myr.
This is substantially shorter than the travel time for warm ionised gas moving at \(\sim600~{\rm km~s^{-1}}\) \citep{shopbell98} to reach \(\sim1\) kpc above the disk, which is \(\sim1.6\) Myr.
Thus, if the observed filaments are advected from the inner wind, they would have experienced several cloud-crushing times during their travel, making their apparent morphological persistence non-trivial \citep{scannapieco15, schneider17}.

Modern simulations suggest that radiative cooling in the turbulent mixing layer at the cloud--wind interface provides the necessary stabilisation: if the cooling time in the mixed gas is shorter than $t_{\rm cc}$, for plausible wind conditions, clouds above a critical size $r_{\rm cl}^{\rm crit}$ of a few pc can survive and even grow by condensing hot wind material \citep{gronke18, sparre19,gronke20}. 
This range is consistent with our observed median width of $\simeq$5.3~pc, suggesting that the detected filaments lie near the survival threshold and possibly are radiatively stabilised against disruption. 
In this picture, the width distribution appears constant with $z$ because structures below $r_{\rm cl}^{\rm crit}$ are rapidly destroyed at all heights, leaving only the surviving population visible at every distance. The lack of inclination-angle evolution further supports this interpretation: if KH instabilities and ram-pressure acceleration had been operating efficiently, one might expect filaments to show progressively more flow-aligned orientations at larger $z$, which is not observed. 
However, the apparent morphological stability should not be interpreted uniquely as evidence for long-lived individual filaments. Continuous formation, projection effects, and surface-brightness selection may also contribute.
\citet{lopez25} estimated cloud cooling times in the M82 wind using a density approximation that assumes a unity covering fraction, finding that the cooling time is shorter than the cloud-crushing time and thus favourable for radiative cooling within the clouds.

Recent M82-specific simulations provide a useful theoretical context for our filament measurements. \citet{li25} and \citet{wang26} find that cool gas is lifted from the central disk, survives early feedback, becomes entrained in the wind, and is stretched into filamentary structures, rather than being produced primarily by net condensation from the hot phase. This supports an interpretation in which the observed \ha\ filaments are surviving or continuously replenished warm structures embedded in the hot wind; however, the lower simulated velocities indicate that additional acceleration or more strongly clustered feedback may still be needed to fully match M82.

\subsection{Implications for wind energetics}

Our line-ratio diagnostics (Section~\ref{subsec_excitation}) indicate that, over the heights probed here, the warm ionized emission in the M82 wind lies at an intermediate excitation state between photoionization and shock excitation. 
Our measurements also suggest a systematic change with height: the diffuse component shows increasingly elevated \sii/\ha\ and \nii/\ha\ at larger distances, whereas the filamentary component remains offset to lower \sii/\ha\ and \nii/\ha, more consistent with photoionized emission from coherent structures. This behaviour suggests that any non-photoionization contribution (e.g. shocks) may become relatively more important at large heights and may preferentially affect the diffuse emission. 
This is consistent with previous long-slit spectroscopic observations of M82 in which inner regions show more star-formation-like excitation while outer regions shift toward shock-like excitation \citep{armus89}. 
However, most previous surveys lack the angular resolution to separate thin filaments from the diffuse background, so this filament--diffuse contrast has not generally been accessible. 
In our case, filaments are selected in \ha\ emission and therefore naturally highlight bright, high-contrast photoionized interfaces; if shocks are present primarily in lower-surface-brightness mixing layers or in unresolved narrow shock fronts, their signature could be diluted in \ha\ and appear more clearly in the diffuse component or on spatial scales smaller than the detected filaments.

In Section~\ref{subsec_excitation}, we have shown that the Lyman-continuum radiation leaking out of the central starburst can readily power the observed emission-line luminosity of the outflow.
If, instead, the optical line emission were predominantly powered by shock heating from the hot wind, the implied energy requirements would be far more severe. In the slow shocks implied by the model fits, the total cooling shock luminosity is typically $\sim$ 50 to 80 $\times$ $L_{\mathrm{H}\alpha}$ \citep{allen08}. Thus, the observed \ha\ luminosity would correspond to a required shock power of \(\sim 5 - 8\times10^{42}\) erg s\(^{-1}\). 
This exceeds the total rate at which kinetic energy is injected by supernovae and stellar winds in M82 \citep[$\sim 3 \times 10^{42}$ erg s$^{-1}$,][]{xu23}. 
In such a scenario, all of the mechanical energy available to drive the wind would be radiated away, leaving insufficient energy to maintain a powerful large-scale outflow. In other words, a shock-dominated interpretation would imply a catastrophically inefficient, or even failed, wind. It is therefore significant that our data favour photoionization, rather than shocks, as the dominant excitation mechanism for the emission-line gas.

This conclusion is broadly consistent with previous work arguing that shocks are energetically inefficient sources of ionizing photons compared to young massive stars.
\citet{gallagher19} show that the majority of outflow components in MANGA galaxies are consistent with photoionization.
They further demonstrate an energetic shortfall: even under extreme assumptions, shocks would predict \ha\ luminosities that are typically one to two orders of magnitude below those observed in bright outflows, making shocks an unlikely dominant power source for the nebular emission. 
Recent analyses reach a similar conclusion using independent datasets \citep{ong25}.

\section{Conclusions}
\label{sec_conclusions}

In this paper, we analyse narrow-band HST imaging of the southern (approaching) outflow of M82 in \ha, \oiii, \sii, and \nii. We detect extraplanar \ha\ emission out to $z\sim$2.1~kpc above the disk, while the fainter lines are traced to smaller heights (with \oiii\ detected out to $z\sim$1.5~kpc). 
Using these data, we characterise the morphology and characteristic scales of the warm ionized structures carried by the starburst-driven wind, separate filamentary and diffuse components, and investigate excitation conditions using multi-line diagnostics. Our main results are as follows:

\begin{enumerate}

\item \textbf{Filament detection and morphology.}
We developed a filament-detection procedure tailored to deep \ha\ imaging, designed to suppress large-scale diffuse emission and isolate narrow coherent structures. The inner wind is highly filamentary, forming a complex interconnected network of strands embedded in a more extended diffuse background.

\item \textbf{Filament properties.}
For each detected filament we measure projected length and width using a non-parametric approach based on the filament mask and distance transform. The filament population has typical dimensions of length $\simeq$9.5~pc and width $\simeq$5.3~pc (medians). The width shows only a weak dependence on length: longer filaments are, on average, slightly broader, but the scatter remains large.
The filament width, length, and inclination angle show no systematic evolution with radial distance above the disk.

\item \textbf{Vertical trends.}
Surface-brightness profiles in all four lines decline monotonically with height. Filaments are locally brighter than the surrounding diffuse emission, but they contribute only a minority of the integrated line flux. The filament flux fraction decreases systematically with height, so that the diffuse component increasingly dominates the optical line luminosity with increasing distances.

\item \textbf{Excitation mechanisms.}
The optical line ratios show that the warm ionized gas occupies an intermediate regime between pure photoionization and shock excitation. The filamentary component remains closer to the photoionization models, consistent with dense cloud surfaces illuminated by ionizing photons leaking from the central starburst. The diffuse component, by contrast, shows systematically enhanced \oiii/\ha, \sii/\ha, and \nii/\ha\ relative to the filaments, and shifts towards the shock-model grids with increasing height. This indicates that additional processes, such as shocks or other non-photoionization heating, become increasingly important in the diffuse gas at larger distances from the disk. The persistent separation between the filamentary and diffuse components demonstrates that the warm ionized wind contains distinct excitation regimes whose relative importance evolves with height.

\item \textbf{Energetics.}
Photoionization by the central starburst is energetically capable of powering the observed \ha\ luminosity of the outflow. The required ionizing photon rate corresponds to only $\sim32\%$ of the photons initially emitted into the outflow bicones. By contrast, a shock-dominated origin for the optical emission would be energetically difficult, because the associated radiative losses would consume all the mechanical power supplied by supernovae and stellar winds. 
This supports a picture in which photoionization provides a major, and likely primary, contribution to the wind \ha\ emission, particularly in the inner bright region and the filaments, while shocks or related non-photoionization processes may contribute secondarily, especially in the diffuse outer wind.

\end{enumerate}

Overall, our results demonstrate that deep, high-resolution narrow-band imaging can directly resolve the parsec-scale structure of the warm ionized phase in a nearby starburst wind, and that separating filamentary and diffuse emission provides new leverage on both the morphology and excitation of multiphase outflows.



\facilities{HST (WFC3)}


\bibliography{main}{}
\bibliographystyle{aasjournalv7}

\end{document}